\documentclass[aps,pra,groupedaddress,superscriptaddress,amssymb,amsmath,showpacs,twocolumn]{revtex4-1}
\usepackage{graphicx}

\begin{document}
 \title{Lasing and high temperature phase transitions in atomic systems with dressed state polaritons}

\author{I. Yu. Chestnov}
\affiliation{Department of Physics and Applied Mathematics, Vladimir State University named after A. G. and N. G. Stoletovs, Gorky str. 87, 600000, Vladimir, Russia}
\author{A. P. Alodjants}
\email[Electronic address]{:alodjants@vlsu.ru}
\affiliation{Department of Physics and Applied Mathematics, Vladimir State University named after A. G. and N. G. Stoletovs, Gorky str. 87, 600000, Vladimir, Russia}
\affiliation{Russian Quantum Center, Novaya str. 100, 143025, Skolkovo, Moscow, Russia}
\author{S. M. Arakelian}
\affiliation{Department of Physics and Applied Mathematics, Vladimir State University named after A. G. and N. G. Stoletovs, Gorky str. 87, 600000, Vladimir, Russia}

\begin{abstract}
We consider the fundamental problem of high temperature phase transitions in the system of high density two-level atoms off-resonantly interacting with a pump field in the presence of optical collisions (OCs) and  placed in the cavity. OCs are considered in the framework of thermalization of atomic dressed state (DS) population. For the case of a strong atom-field coupling condition we analyze the problem of thermodynamically equilibrium superradiant phase transition for the order parameter representing a real amplitude of cavity mode and taking place as a result of atomic DSs thermalization process. Such transition is also connected with condensed (coherent) properties of low branch (LB) DS-polaritons occurring in the cavity. For describing non-equilibrium phase transitions we derive Maxwell-Bloch like equations which account for cavity decay rate, collisional decay rate and spontaneous emission. Various aspects of transitions to laser field formation by using atomic DS levels for both positive and negative detuning of a pump field from atomic transition frequency are studied in detail.  It is revealed, that for positive atom-light detuning DS lasing can be obtained in the presence of quasi-equilibrium DS population that corresponds to a true two-level atomic system with the inversion in nonresonant limit.
\end{abstract}
\maketitle
\section{INTRODUCTION}

Lasing and Bose-Einstein condensation (BEC) are two phenomena which make photonic and matter waves macroscopically coherent when they are confined in the cavity and trapped respectively, see e.g. \cite{1,2,3}. Although the final state of the ensemble of particles is spontaneously broken, and can be described by well-defined amplitude and phase, the physical reasons for achieving this state are completely different for lasers and for objects exhibiting BEC, cf. \cite{4}.

Nowadays lasing is recognized as a very general and universal phenomenon that can be obtained under the matter-light field interaction in the cavity containing different media, see e.g. \cite{5}. Lasers are non-equilibrium systems in the sense of thermodynamics and, in the common case, require pumping in order to form a population inversion, which elucidates non-equilibrium features of the lasing process. In many cases, such inversion can be reached in the medium with a three (or more) energy level configuration only. Steady state population inversion cannot be achieved for a two-level system interacting with e.m. field in the cavity in the context of a semiclassical treatment, because probabilities of pump-induced upward and downward transitions between these two levels are equal. However, as it is shown in \cite{6a} and recently discussed in \cite{6b}, a full quantum theory of atom-light interaction predicts the presence of such inversion at the steady state.

The physical picture  becomes richer in case when coherent light field formation takes place due to the equilibrium (or quasi-equilibrium) phase transition occurring in a coupled matter-light system. In fact, population inversion does not play any role in this case because it does not correspond to equilibrium properties of the system and it is not relevant for such transitions. Actually, that is why a two-level system is suitable for realizing a thermodynamically equilibrium phase transition under the matter-light interaction, cf. \cite{31}. Noticing that dissipation and decoherence effects must be maximally suppressed in this case. If such effects are not so weak a coupled matter-field system undergoes non-equilibrium transition to lasing -- see \cite{22}. However, this transition behaves quite differently as compared to conventional lasers because it occurs in the inversionless two-level system coupled to quantized field irradiation and strongly depends on the specific features of the medium, cf. \cite{21}.

Nowadays it has been shown that different condensed matter, solid state and even photonic systems exhibit BEC phenomenon firstly obtained with ultra-cold atoms, see e.g.  \cite{6,7,8,9,10}. Although not all of them are completely thermodynamically equilibrium, in practice, the requirement of thermalization plays the essential role in achieving BEC state, cf. \cite{2}.

Recently, BEC phenomenon has been observed for low dispersion branch exciton-polaritons in high-finesse semiconductor microcavities, see e.g. \cite{Deng}. Since polaritons are mixed light-matter states, such experiments allow to resume the old discussion about physical similarities between lasing and phase transition at a qualitatively new level, cf.  \cite{3,4,11}. Moreover, the terms ``polariton laser'' \cite{22} or ``atom laser'' \cite{Bloch} which appeared in the last decades for emphasizing experimentally feasible high coherent properties of matter waves attach practical value to the discussion. Actually, two thresholds are clearly seen in the experiments with exciton-polaritons confined in the cavity. The first one occurs at low pumping intensities and characterizes inversionless systems with excitons strongly coupled with a cavity mode \cite{12}. This is the case of ``polariton laser'' that corresponds to quasi-equilibrium state of the system, cf. \cite{22}. By increasing the intensity of a pumping field it is possible to achieve the second threshold that corresponds to photon lasing.

The situation becomes physically intriguing in the atomic physics area operating with coupled atom-light states.  Keystone problem of observing of atomic polariton BEC is connected with the problem of achieving true thermal equilibrium (or quasi-equilibrium) for coupled atom-field states \cite{14}. Great interest in thermodynamic properties of atomic gases in the presence of optical field irradiation has been evoked recently \cite{15, 16}. Dressed states (DSs) are at the heart of describing the interaction of  high density atomic gases with nonresonant radiation in the presence of collisions with buffer gas particles, so-called  optical collisions (OCs) \cite{17, 18}. Previously we suggested a model of atomic DS thermalization being obtained due to OC processes that accounts for the evolution of pseudo-spin Bloch vector components and characterizes the essential (negative) role of a spontaneous emission in the thermalization process \cite{19}.  Recently, we have shown that in the presence of both photon and polariton trapping in
biconical waveguide cavity with the appropriate lifetime of polaritons a high temperature BEC for low branch (LB) photon-like polaritons is achieved \cite{20}.

Some time ago non-equilibrium phenomena in DS picture were intensively discussed  in the framework of lasing phenomena in inversionless medium; these phenomena taking place for lasers which operate on a transition between DS levels, or simply for DS lasers -- see \cite{23,24,25,26,Berg}. The systems considered for these purposes imply the usage of spontaneous processes playing the role of pumping. However, OC processes which we deal with in the paper can represent a much more effective mechanism for population redistribution of DSs. In particular, OC-induced transitions between DS are not identical for upward and backward directions. That is why lasing can be obtained in a two-level DS system in the presence of OCs, cf. \cite{16}.

The main objective of the paper is to study both equilibrium and non-equilibrium phase transitions occurring in the atomic system with two generic levels being under OCs with high pressure buffer gas particles and interacting with the field in the cavity with finite Q-factor.

The paper is arranged as follows. In Sec. II we discuss some equilibrium and non-equilibrium schemes and methods of the atom-field interaction involving DS transitions and lasing resulting in the origin of the coherent field state in a two-level atomic system without cavity. Some principal experimental results obtained previously are discussed in this section. In Sec. III we develop the theory of cavity quantum electrodynamics (QED) with DSs deriving Maxwell-Bloch-like equations for atomic DS polarization, population inversion and cavity field for a high density ensemble of two-level atoms being under OCs with buffer gas atoms in the presence of a pump field and interacting with cavity mode simultaneously. In Sec. IV we examine the effect of thermodynamically equilibrium phase transition to the superradiant state occurring in such a system. Various aspects of non-equilibrium phase transitions, temporal dynamics of DS variables and DS lasing phenomenon  occurring in the cavity with finite  Q-factor  in the presence of  collisions with high pressure buffer gas particles are studied in details  in Sec. V and Sec.VI, respectively. In conclusion, we summarize the results obtained.

\section{LASING WITH DS}

Let us consider a high density two-level atomic system non-resonantly interacting with a quantized light field without the cavity. Basic description of the interaction of atoms with optical field can be done by using DS defined as,  cf. \cite{17},
\begin{subequations} \label{eq1_}
\begin{eqnarray}
        {\left| 1(N) \right\rangle} =\sin \theta {\left| a,N+1 \right\rangle} +\cos \theta {\left| b,N \right\rangle}, \\
        {\left| 2(N) \right\rangle} =\cos \theta {\left| a,N+1 \right\rangle} -\sin \theta {\left| b,N \right\rangle},
\end{eqnarray}
\end{subequations}
where ${\left| 1(N) \right\rangle} $ and ${\left| 2(N) \right\rangle} $ are upper and lower levels for DSs containing the number of photons $N$. The states ${\left| a,N+1 \right\rangle} $ and ${\left| b,N \right\rangle} $ correspond to uncoupled (bare) atom-light states, respectively. The coefficients
\begin{subequations} \label{eq2_}
\begin{equation*}
\sin \theta =\frac{1}{\sqrt{2} } \sqrt{1+\frac{\delta }{\Omega _{R} } }, \ \ \cos \theta =\frac{1}{\sqrt{2} } \sqrt{1-\frac{\delta }{\Omega _{R} } } \eqno{(\ref{eq2_}\rm{a,b})}
\end{equation*}
\end{subequations}
\noindent determine the contribution of bare states to DS levels ${\left| 1(N) \right\rangle} $ and ${\left| 2(N) \right\rangle} $; $\Omega _{R} =\sqrt{\delta ^{2} +\Omega^{2} } $ is generalized Rabi frequency, $\delta =\omega _{L} -\omega _{\rm{at}} $ is atom-light detuning and $\Omega$ represents resonant ($\delta=0$) Rabi frequency that generally depends on photon number $N$ in the pump field \cite{17}.

In the paper we focus on the non-resonant (or so-called perturbative) limit when inequality
\begin{equation} \label{eq3_}
|\delta | \gg \Omega
\end{equation}
is fulfilled for large enough $N$, i.e. $N \gg 1$.

\begin{figure}
\includegraphics[scale=0.43]{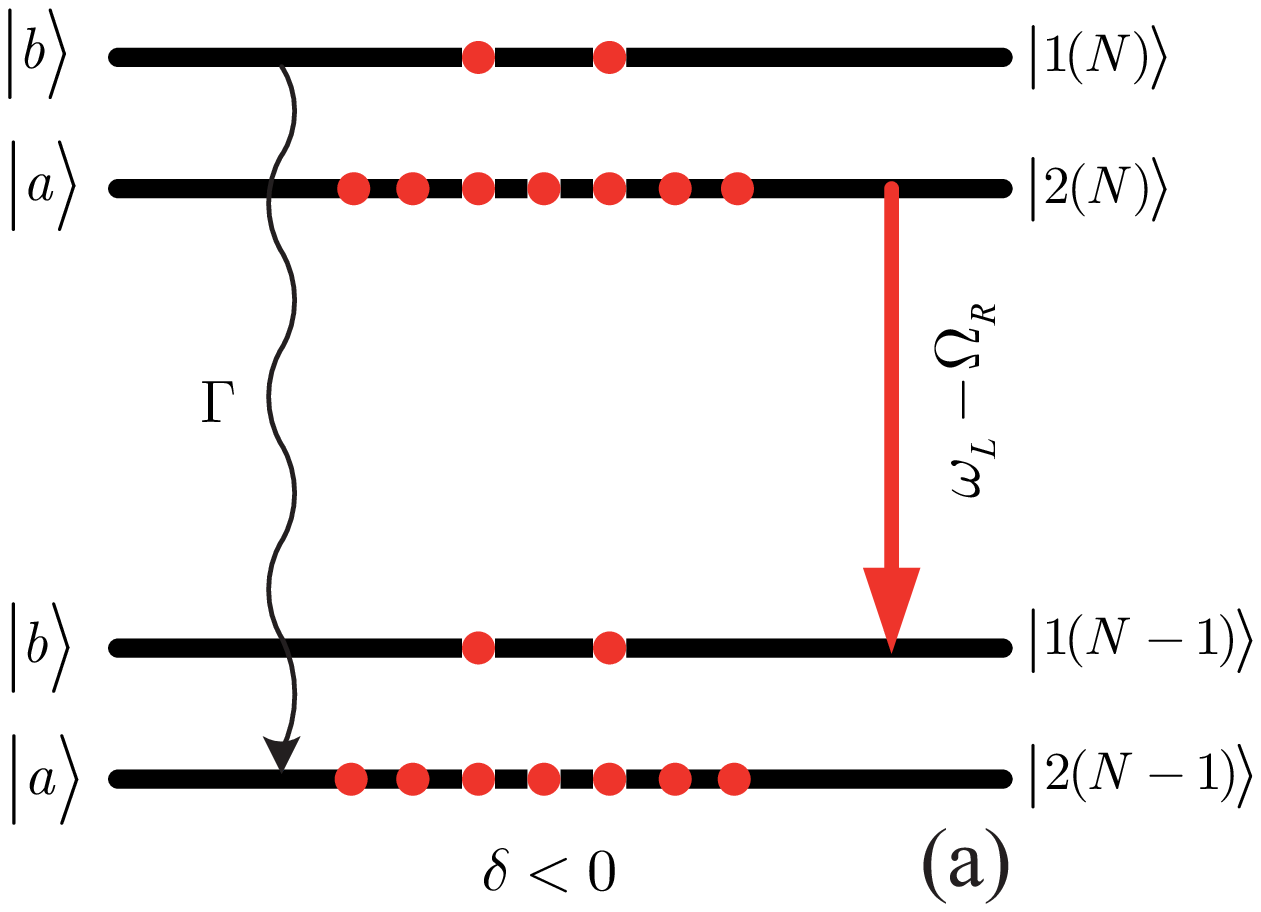}
\includegraphics[scale=0.43]{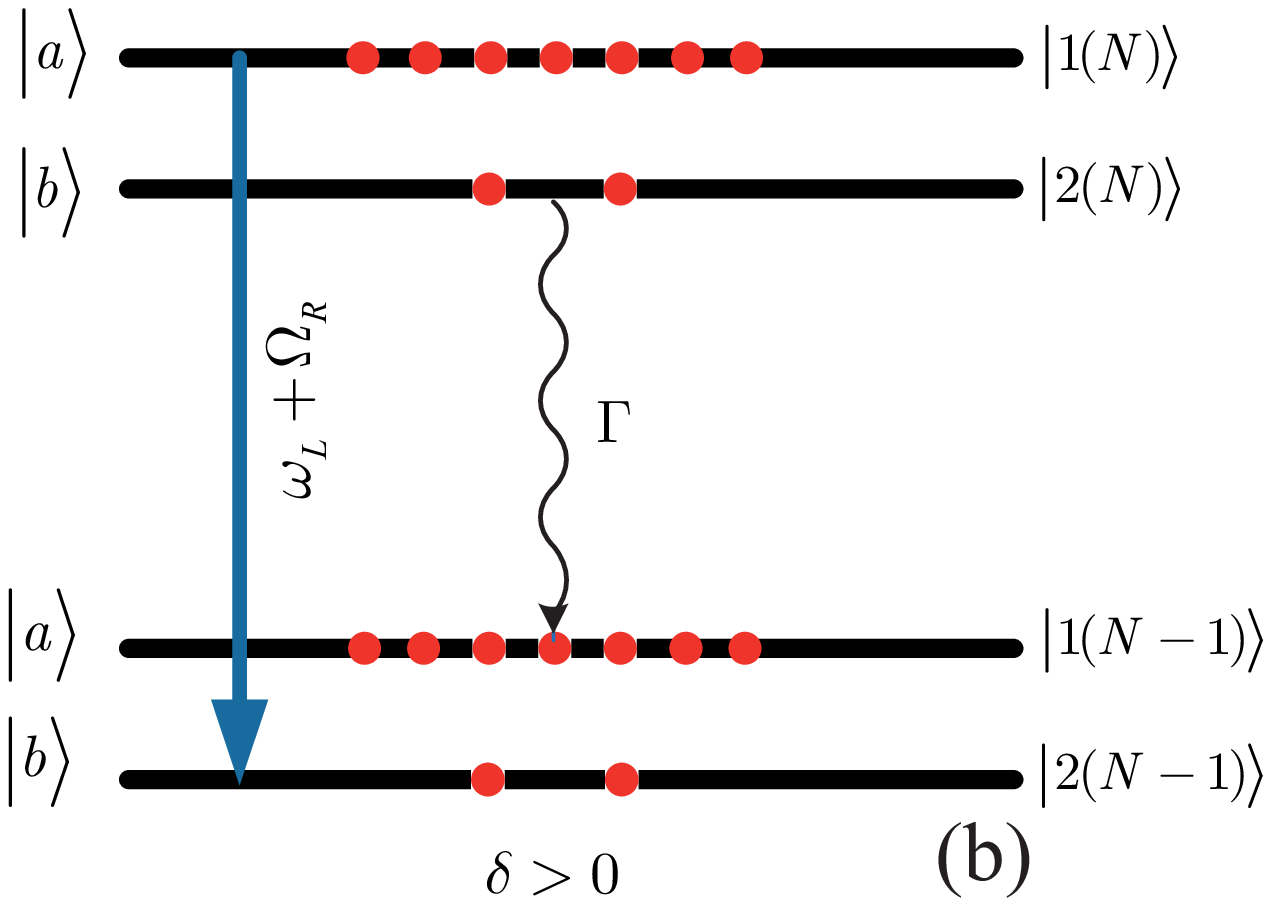}
\includegraphics[scale=0.43]{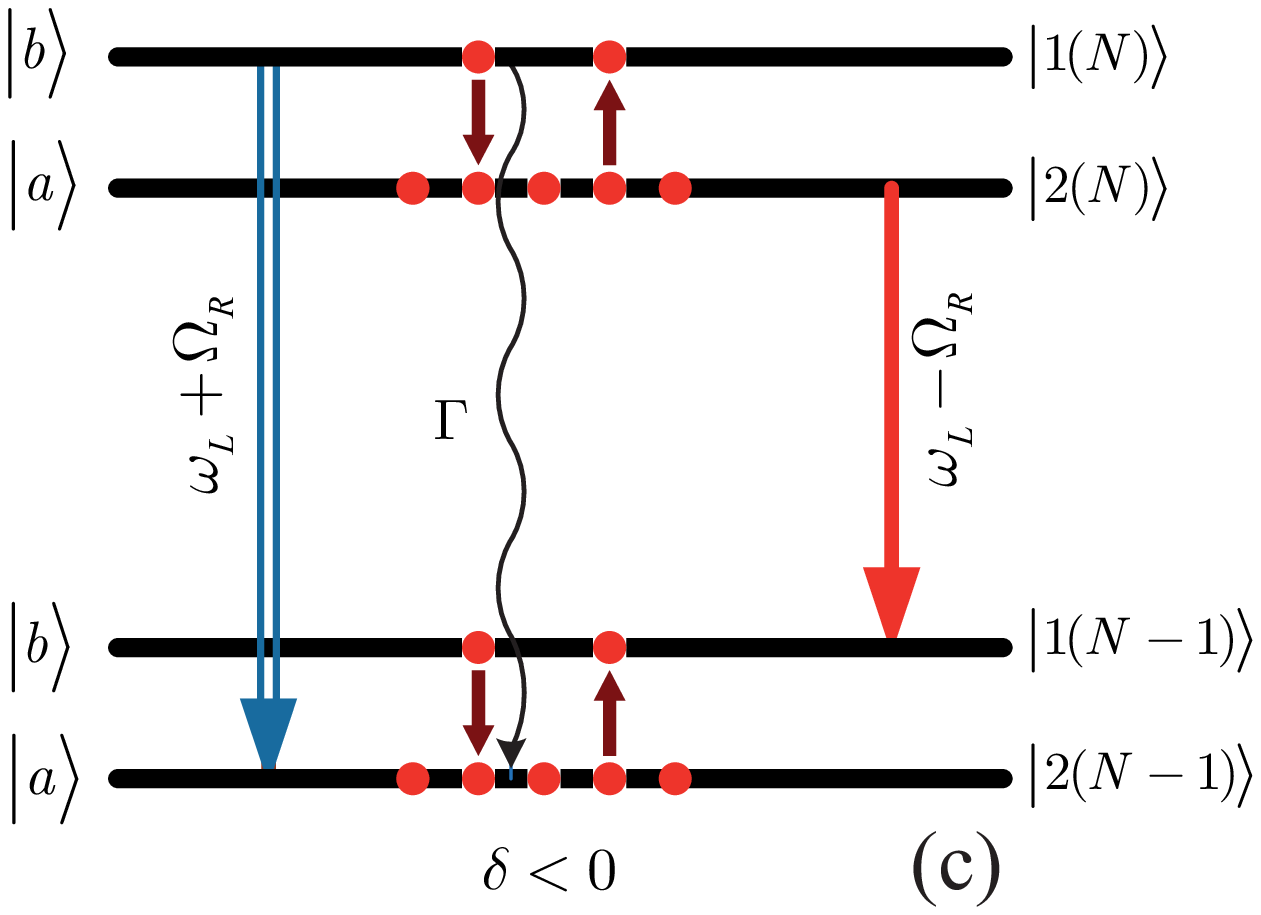}
\includegraphics[scale=0.43]{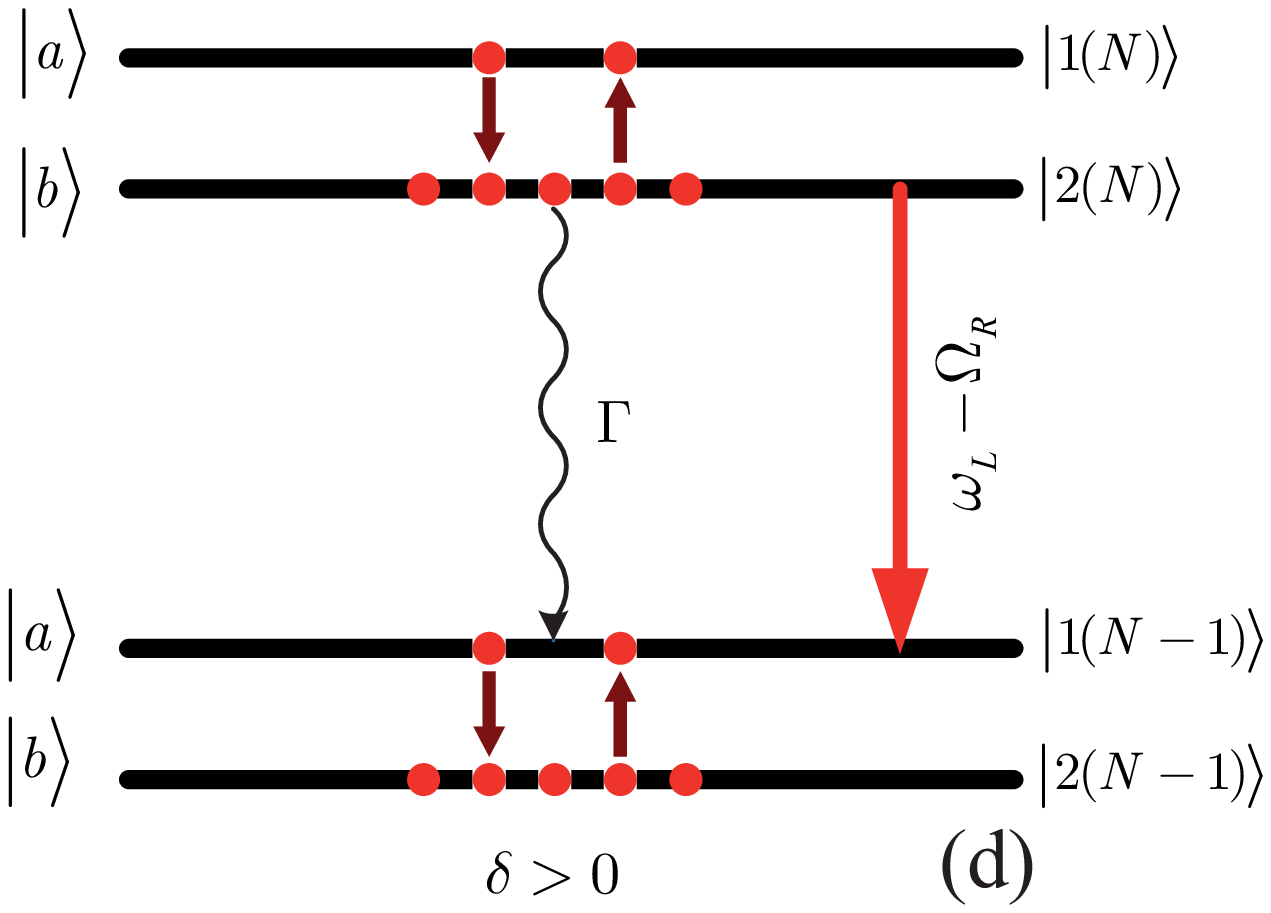}
\caption{\label{fig1} (Color online) Energy level manifolds in the coupled atom-light system without (a,b) and with (c,d) DS thermalization under the perturbative condition \eqref{eq3_} for (a,c) -- negative and (b,d) -- positive atom-light detunings. Bold vertical lines indicate the transition involved in DS laser field generation. Wavy line corresponds to spontaneous emission transitions between DSs being responsible for DS population redistribution in the absence of thermalization. Double line on (c) indicates equilibrium  phase transition to superradiant state.}
\end{figure}

In Fig.\ref{fig1} we represent two adjacent manifolds of DS levels for the different sign of atom-light detuning $\delta $ under perturbative limit. Optically allowed transitions between these states form the so-called Mollow triplet, that consists of the central component with frequency $\omega _{L} $ and two sidebands shifted onto $\pm \Omega _{R}$.

In the absence of pressure broadening and accompanying OC processes, the ground atomic state  ${\left| a \right\rangle} $ is much more populated due to the spontaneous emission. This leads to the establishment of population inversion between DS levels in a considerably non-resonant region when condition \eqref{eq3_} is fulfilled. It is worth noticing, that the amplification in the system of  two-level atoms off-resonantly interacting with a high intensity field has been discussed \cite{27a} and later experimentally demonstrated in \cite{27}. Actually, under condition \eqref{eq3_}, for $\delta <0$ (see Fig.\ref{fig1}(a)) from \eqref{eq1_}, \eqref{eq2_}, we have $\sin \theta \approx 0$ and $\cos \theta \approx 1 $, and a lower DS ${\left| 2(N) \right\rangle} $ corresponds to the bare state ${\left| a,N+1 \right\rangle} $ which describes the atoms in the ground state. As a result population inversion is established between the states ${\left| 2(N) \right\rangle} $ and ${\left| 1(N-1) \right\rangle} $; this giving a possibility to lasing and amplification of the red fluorescence component $\omega _{L} -\Omega _{R} $. Previously, laser field generation on this transition has been obtained in \cite{26} by using sodium atoms in a low pressure (less than $4\cdot 10^{-4} $ bar) buffer gas environment and in \cite{24} by exploring beam of barium atoms.

For positive detuning $\delta >0$  (see Fig.\ref{fig1}(b)), it is easy to show, that there is a population inversion between levels ${\left| 1(N) \right\rangle} $ and ${\left| 2(N-1) \right\rangle} $, that corresponds to the blue wing of Mollow triplet with frequency $\omega _{L} +\Omega _{R} $. The generation on this transition has been also observed, see e.g. \cite{23} and later discussed in \cite{25}. This transition can be also recognized as Raman process that involves transition ${\left| a \right\rangle} \to {\left| b \right\rangle} $ accompanied by the emission of photon from the ground to the excited atomic state, cf. \cite{28}. This process is also called Rabi- or three photon gain process \cite{29}.  Noticing that in both cases in Fig.\ref{fig1}(a),(b) laser generation occurs in the inversionless ``two-level'' atomic system.

In the presence of OCs, collisionally aided excitation allows a transfer between DS components ${\left| 1(N) \right\rangle} $ and ${\left| 2(N) \right\rangle} $. If such processes are fast enough, DS population distribution approaches its equilibrium state characterized by Boltzmann distribution -- see \cite{19}. As a result, in a completely thermalized system the population of ${\left| 2(N) \right\rangle} $ should be larger than that of the upper one by Boltzmann factor $\exp \left[\frac{\hbar \Omega _{R} }{k_{B} T} \right]\approx \exp \left[\frac{\hbar \left|\delta \right|}{k_{B} T} \right]$, see e.g. Fig. \ref{fig1}(c), (d). Particularly, in the limit of large and negative $\delta$ characterizing the inversionless two-level atomic system, the equilibrium distribution established between upper and lower DS levels enables us to expect a thermodynamically equilibrium phase transition to the superradiant state (see double line in Fig. \ref{fig1}(c)).

 A physical picture changes dramatically for positive detuning $\delta >0$ -- see Fig.\ref{fig1}(d).  For the large positive atom-light field detuning a lower DS ${\left| 2(N) \right\rangle} $ corresponds to the excited atomic level ${\left| b,N \right\rangle} $ which is much more populated now. In other words, we achieve the inversion in a true \emph{two-level} atomic system in the presence of thermodynamically equilibrium or quasi-equilibrium state, cf. \cite{6a, 16}. In fact, a thermal buffer gas reservoir acts as a pumping for achieving population inversion. Apart from the case discussed above (see Fig.\ref{fig1}(b)) lasing in the system can be obtained in ${\left| 2(N) \right\rangle} \to {\left| 1(N-1) \right\rangle} $ transition -- see Fig.\ref{fig1}(d), that corresponds approximately to transition ${\left| b \right\rangle} \to {\left| a \right\rangle} $ in terms of real atomic states under the condition \eqref{eq3_}.

In \cite{16} laser field gain has been demonstrated experimentally under the excitation of the blue wing of fluorescence intensity component of sodium atoms being at temperatures T=600 K under the pressure from 600 torrs to 4 atmospheres of helium buffer gas and interacting with the optical field in a single pass regime. Such conditions did not allow to obtain a thermodynamically equilibrium state for DS population, cf. \cite{19}. However, they were sufficient to reach the population inversion for true atomic levels.

\section{DS CAVITY QED}

\begin{figure*}
\centering
\includegraphics[scale=0.5]{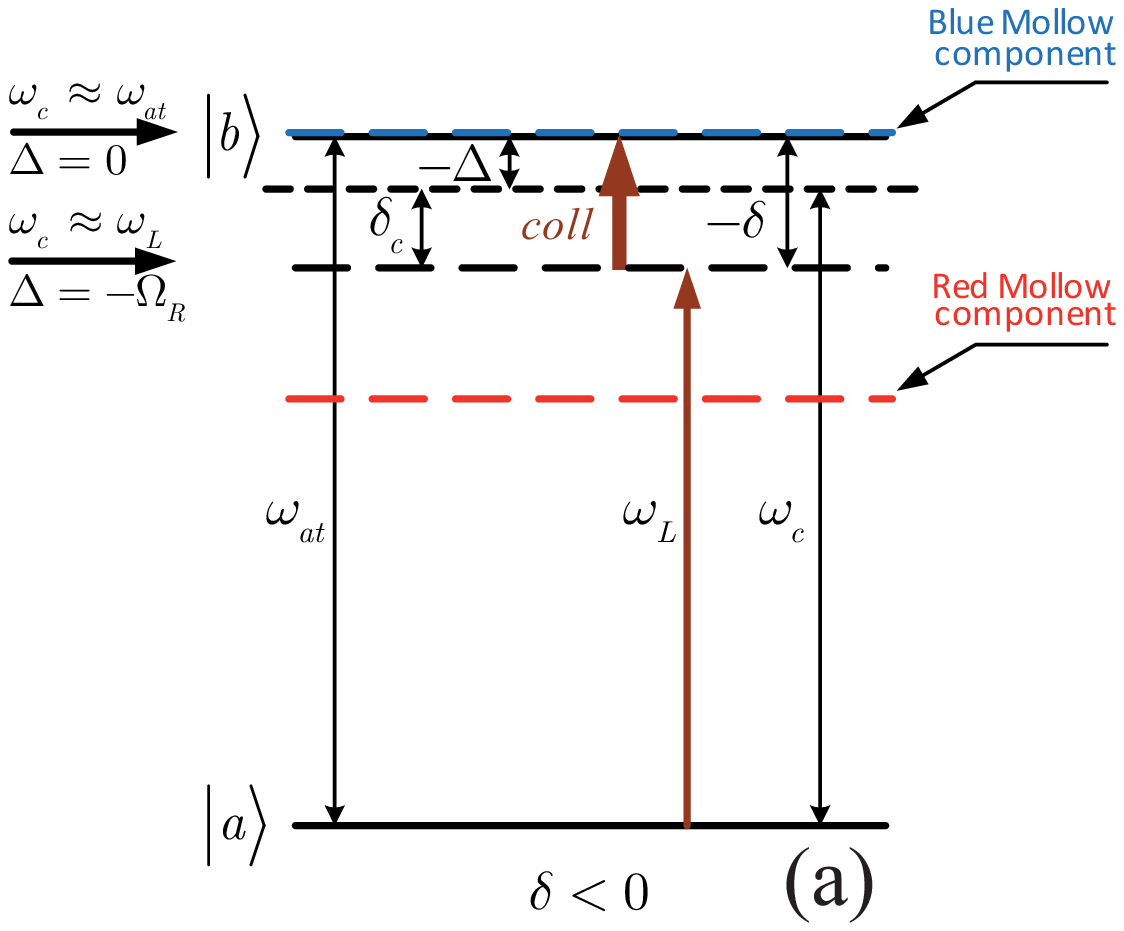}
\includegraphics[scale=0.5]{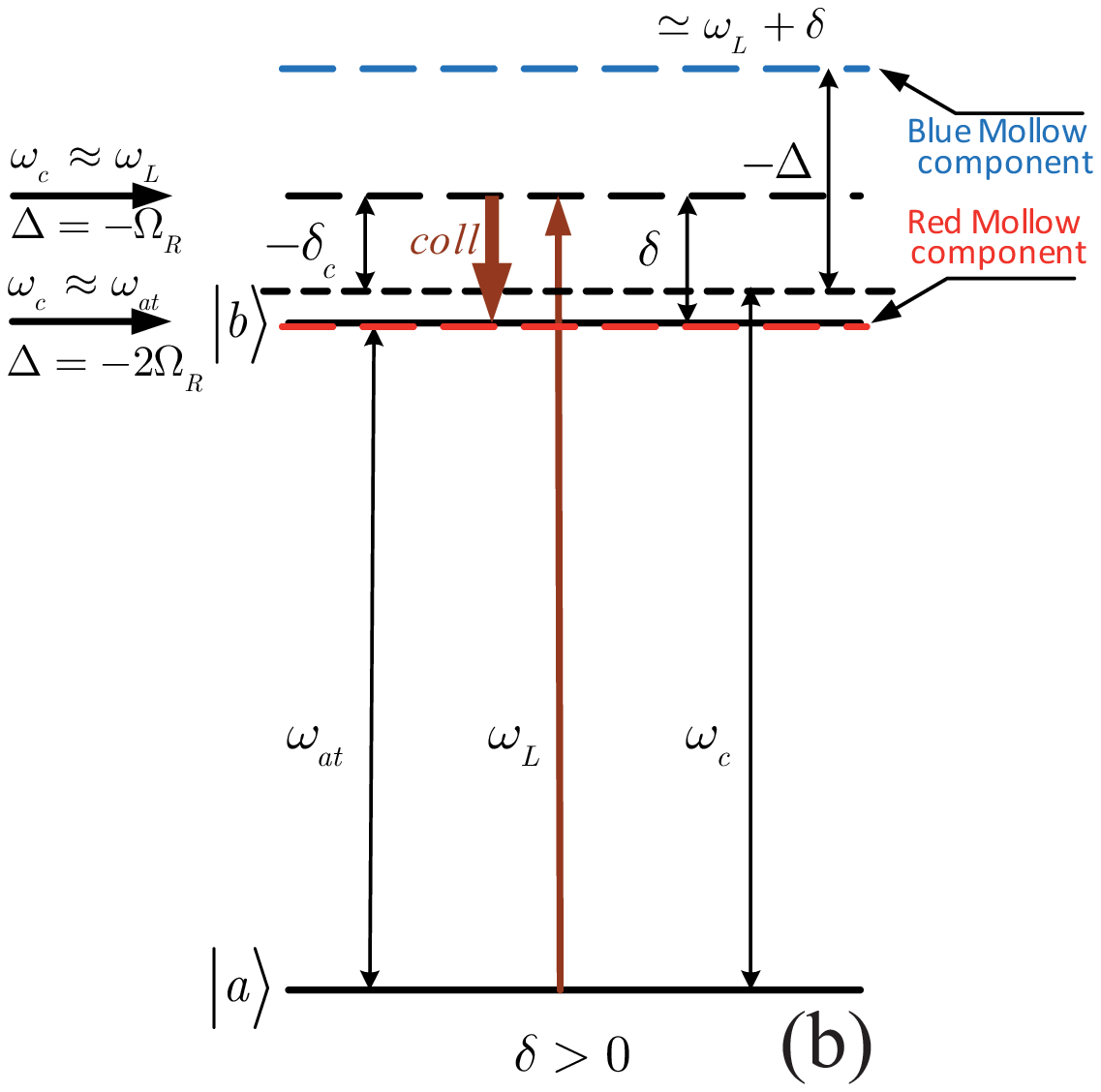}
\caption{\label{fig2} (Color online) Schematic representation of nonresonant absorption of pump field with frequency $\omega _{L} $ for (a) negative atom-field detuning, $\delta \equiv \omega _{L} -\omega _{\rm{at}} <0$, and (b) positive atom-field detuning, that is, $\delta >0 $ in the perturbative limit  \eqref{eq3_}; $\omega _{c} $ is  frequency  of  cavity field,  with relevant detuning $\Delta \simeq \omega _{c} -\omega _{L} -\left|\delta \right|$. For positive detuning $\delta >0$, in the presence of OCs the atom has some energy excess $\hbar \delta $, which is transformed into to the kinetic energy of the atoms after the collision.}
\end{figure*}

Let us consider the case when the ensemble of  $N_{\rm{at}} $ two-level atoms interacts with the optical field in the presence of OCs in the cavity. Total Hamiltonian $H=H_{\rm{ATF}} +H_{\rm{C}} $ can be represented as a sum of part $H_{\rm{ATF}} $ that characterizes the interaction of two-level atoms with a pump light field; $H_{\rm{C}} $ corresponds to the interaction of  atoms with a cavity mode tuned to  frequency $\omega _{c} $ and described by annihilation (creation) operators $a$ ($a^{\dag } $). We represent $H_{\rm{ATF}} $ and $H_{\rm{C}} $ in the form
\begin{widetext}
\begin{subequations} \label{eq5_}
\begin{eqnarray}
H_{\rm{ATF}} =\hbar \omega _{L} f^{\dag} f+\hbar \sum _{j}^{N_{\rm{at}} }\left(\frac{\omega _{\rm{at}}}{2} \left(\left| b \right\rangle_{j} {{\vphantom{left(\left. \right\rangle}}_{j}\left\langle b \right|}  - {\left| a \right\rangle} _{j} {{\vphantom{left(\left. \right\rangle}}_{j}\left\langle a \right|} \right) +g\left(s_{+j} f+s_{-j} f^{\dag } \right)\right),\\
H_{\rm{C}} =\hbar \omega _{c} a^{\dag } a+\frac{\hbar \kappa }{\sqrt{N_{\rm{at}} } } \sum _{j}^{N_{\rm{at}} }\left(s_{+j} a+s_{-j} a^{\dag } \right),
\end{eqnarray}
\end{subequations}
\end{widetext}
where $f$ ($f^{\dag } $) is annihilation (creation) operator for the photons absorbed (or emitted) under the interaction with the pumping field having frequency $\omega _{L} $ -- see Fig.\ref{fig2}, $g=\left({\left|d_{ab} \right|^{2} \omega _{L} \mathord{\left/ {\vphantom {\left|d_{ab} \right|^{2} \omega _{L}  2\hbar \varepsilon _{0} V}} \right. \kern-\nulldelimiterspace} 2\hbar \varepsilon _{0} V} \right)^{1/2} $ is atom-field interaction constant, which are supposed to be identical for all $N_{\rm{at}} $ atoms, $d_{ab} $ is  atomic dipole matrix element, and $V$ is  atom-field interaction volume in the cavity, $\kappa =\left({\left|d_{ab} \right|^{2} N_{\rm{at}} \omega _{c} \mathord{\left/ {\vphantom {\left|d_{ab} \right|^{2} \omega _{L}  2\hbar \varepsilon _{0} V}} \right. \kern-\nulldelimiterspace} 2\hbar \varepsilon _{0} V} \right)^{1/2} $ is  cooperative parameter of the interaction of cavity field with $N_{\rm{at}} $ two-level atoms; $s_{-j} ={\left| a \right\rangle} _{j} {{\vphantom{left(\left. \right\rangle}}_{j}\left\langle b \right|} $ and $s_{+j} =s_{-j}^{\dag } ={\left| b \right\rangle} _{j} {{\vphantom{left(\left. \right\rangle}}_{j}\left\langle a \right|} $ represent atomic transition operators for the $j$-th atom.

Then, the estimate of Eqs. \eqref{eq5_} presumes transition to DS basis, cf. \cite{25}. Considering a pump field as a classical one that possesses the average photon number $N_{0}$ we suggest to substitute $f^{\dag}f \rightarrow \sqrt{N_{0}}$ in Eq. (\ref{eq5_}a) and assume the so-called semiclassical DSs for further calculations, cf. \cite{Berg}. In this case, the transition to DS basis could be realized by using the unitary operator $U =\prod_{j}\exp\left[i \theta s_{2j} \right]$, where $s_{2j}=\left(s_{+j}-s_{-j}\right)/i$, see \cite{25} for more details.

A more rigorous approach to the transformation of Eqs. \eqref{eq5_} into the DS basis could be achieved with the usage of completness relation $I= \sum _{N=0}^{\infty} \left| N \right\rangle \left\langle N \right|$ in a relevant representation of the Fock state basis for pump field operators, that is, $f^{\dag}f = \sum _{N=0}^{\infty} N \left| N \right\rangle \left\langle N \right|$, $f = \sum _{N=0}^{\infty} \sqrt{N} \left| N-1 \right\rangle \left\langle N \right|= \sum _{N=0}^{\infty} \sqrt{N+1} \left| N \right\rangle \left\langle N+1 \right|$, cf. \cite{35}.
For example, we can rewrite Eq. (\ref{eq5_}a) as
\begin{widetext}
\begin{multline} \label{eq6_}
 H_{\rm{ATF}} =\hbar \sum _{N=0}^{\infty} \left[ \omega_{L} N {\left| N \right\rangle}{\left\langle N \right|}
  + \sum _{j}^{N_{\rm{at}} }\left[\frac{\omega_{\rm{at}}}{2}\left({\left| b,N \right\rangle} _{j} {{\vphantom{left(\left. \right\rangle}}_{j}\left\langle b,N \right|} - {\left| a,N+1 \right\rangle} _{j} {{\vphantom{left(\left. \right\rangle}}_{j}\left\langle a,N+1 \right|}  \right) \right.\right. + \\
 +\left. \left. \frac{\hbar \Omega_{N} }{2} \left({\left| a,N+1 \right\rangle} _{j} {{\vphantom{left(\left. \right\rangle}}_{j}\left\langle b,N \right|}   +{\left| b,N \right\rangle} _{j}  {{\vphantom{left(\left. \right\rangle}}_{j}\left\langle a,N+1 \right|} \right) \right] \right],
 \end{multline}
\end{widetext}
where $\Omega _{N} =2g \sqrt{N+1}$ is photon number dependent resonant Rabi frequency.

However, for large $N \gg 1$ and small photon number fluctuations in the pump field it is justified to assume that in Eq. \eqref{eq6_} $\Omega_{N} \simeq \Omega=2g\sqrt{N_{0}}$ and $\Omega_R \simeq \sqrt{\delta ^2 +4g^2 N_{0}}$, cf. \cite{17, Ast}.

Taking into account the definitions \eqref{eq1_} for DSs  and moving to the rotating frame, that is, relative to frequency $\omega _{L} $, we can  recast total atom-field Hamiltonian $H$ in the DSs basis:
\begin{widetext}
\begin{multline} \label{eq7_}
 H =\hbar \delta _{c} a^{\dag } a+\frac{\hbar \Omega _{R} }{2} \sum _{j}^{N_{\rm{at}} }S_{zj}^{\rm{DS}}  +\frac{\hbar \kappa }{2\sqrt{N_{\rm{at}} } } \sum _{j}^{N_{\rm{at}} }\left[\left\{S_{zj}^{\rm{DS}} \sin 2\theta +2 S_{+j}^{\rm{DS}} \cos ^{2} \theta -2 S_{-j}^{\rm{DS}} \sin ^{2} \theta \right\}a\right. + \\ +\left. \left\{S_{zj}^{\rm{DS}} \sin 2\theta +2S_{-j}^{\rm{DS}} \cos ^{2} \theta -2 S_{+j}^{\rm{DS}} \sin ^{2} \theta\right\}a^{\dag } \right],
 \end{multline}
\end{widetext}
where $\delta _{c} =\omega _{c} -\omega _{L} $ is detuning of the cavity mode from frequency $\omega _{L} $ of a pump field. In \eqref{eq7_} we give definitions
\begin{subequations} \label{eq8_}
\begin{gather*}
S_{-j}^{\rm{DS}} =\sum _{N=0}^{\infty}{\left| 2(N) \right\rangle} _{j} {{\vphantom{left(\left. \right\rangle}}_{j}\left\langle 1(N) \right|}, \tag{\ref{eq8_}\rm{a}}\\
 S_{+j}^{\rm{DS}} =\sum _{N=0}^{\infty}{\left| 1(N) \right\rangle} _{j} {{\vphantom{left(\left. \right\rangle}}_{j}\left\langle 2(N) \right|}  \tag{\ref{eq8_}\rm{b}},\\
S_{zj}^{\rm{DS}} =\sum _{N=0}^{\infty}\left({\left| 1(N) \right\rangle} _{j} {{\vphantom{left(\left. \right\rangle}}_{j}\left\langle 1(N) \right|}  -{\left| 2(N) \right\rangle} _{j} {{\vphantom{left(\left. \right\rangle}}_{j}\left\langle 2(N) \right|} \right), \tag{\ref{eq8_}\rm{c}}
\end{gather*}
\end{subequations}
taking the sum throughout all photon numbers $N$ in DS manifolds, cf. \cite{19}.

In the paper we restrict ourselves by one-photon transitions only, taking into account neighbor manifolds -- see Fig.\ref{fig1}. Hence, we can transfer to rotating wave approximation allocating appropriate processes in DS basis. Transitions between the states ${\left| 1(N) \right\rangle} \to {\left| 2(N-1) \right\rangle} $ (for brevity we call it ${\left| 1 \right\rangle} \to {\left| 2 \right\rangle} $) -- see Fig.\ref{fig1}(b), (c), correspond to the destruction of excitation at the upper dressed level ${\left| 1(N) \right\rangle} $ (operator $S_{-j}^{\rm{DS}}$ in Eq. (\ref{eq8_}a)) with the emission of a cavity photon characterized by operator $a^{\dag}$. As a result Eq. \eqref{eq7_} reduces to:
\begin{multline} \label{eq9_}
 H\equiv H_{12} =\hbar \delta _{c} a^{\dag } a+ \\
 + \sum _{j}^{N_{\rm{at}} } \left[ \frac{\hbar \Omega _{R} }{2} S_{zj}^{\rm{DS}}  +\frac{\hbar \kappa_{12}}{\sqrt{N_{\rm{at}} } } \left(S_{+j}^{\rm{DS}} a+S_{-j}^{\rm{DS}} a^{\dag } \right) \right],
\end{multline}
where we introduce parameter $\kappa_{12}=\kappa \cos ^{2} \theta $ that describes the effective coupling of the atomic DS ensemble with a cavity mode when cavity frequency is tuned close to the frequency of transition ${\left| 1(N) \right\rangle} \to {\left| 2(N-1) \right\rangle} $.

Another type of transitions in the atomic system with DSs can be associated with the process ${\left| 2(N) \right\rangle} \to {\left| 1(N-1) \right\rangle} $ (see Fig.\ref{fig1}(a), (d)) that is relevant to the emission of a cavity photon under the transition from a lower DS level ${\left| 2 \right\rangle} $ to the upper one ${\left| 1 \right\rangle} $. In this limit effective Hamiltonian \eqref{eq7_} reduces to
\begin{multline} \label{eq10_}
H\equiv H_{21} =\hbar \delta _{c} a^{\dag } a+\\
+ \sum _{j}^{N_{\rm{at}} } \left[ \frac{\hbar \Omega _{R} }{2} S_{zj}^{\rm{DS}}  -\frac{\hbar \kappa_{21}}{\sqrt{N_{\rm{at}} } } \left(S_{-j}^{\rm{DS}} a+S_{+j}^{\rm{DS}} a^{\dag } \right) \right],
\end{multline}
where $\kappa_{21}=\kappa \sin ^{2} \theta $ presumes an effective coupling coefficient for transition ${\left| 2(N) \right\rangle} \to {\left| 1(N-1) \right\rangle} $, or ${\left| 2 \right\rangle} \to {\left| 1 \right\rangle} $ for brevity.

In the perturbative limit \eqref{eq3_} the efficiency of atom-cavity field  coupling depends essentially on the sign of atom-pump light detuning $\delta $. In particular, for coupling parameter $\kappa_{12}$ in \eqref{eq9_} we obtain
\begin{subequations} \label{eq11_}
\begin{eqnarray}
\kappa_{12} \approx \kappa \left(1-\frac{\Omega^{2} }{4\delta ^{2} } \right)\to \kappa \ \ \rm{for} \ \ \delta <0, \\
\kappa_{12} \approx \frac{\kappa \Omega^{2} }{4\delta ^{2} } \ll \kappa \ \ \rm{for} \ \ \delta >0.
\end{eqnarray}
\end{subequations}

The Eq. (\ref{eq11_}a) implies that it is easier to realize a strong atom-cavity field coupling condition for which  effective parameter  $\kappa_{12}$ exceeds dissipation, decoherence and/or dephasing  effects for negative detuning $\delta <0$, see \eqref{eq29_} and Fig.\ref{fig1}(a). As it was discussed in the previous section the case of $\delta >0$  corresponds to Raman-type transitions with frequency $\omega _{L} +\Omega _{R} $; this being  relevant to a weak effective coupling between atomic DSs and cavity mode -- see. Fig.\ref{fig1}(b). This limit corresponds to DS laser described in \cite{25}.

For parameter $\kappa_{21}$, given in Eq. \eqref{eq10_}, the situation is opposite, that is,
\begin{subequations} \label{eq12_}
\begin{eqnarray}
\kappa_{21} \approx \frac{\kappa \Omega^{2} }{4\delta ^{2} } \ll \kappa \ \ \rm{for} \ \  \delta <0, \\
\kappa_{21} \approx \kappa \left(1-\frac{\Omega^{2} }{4\delta ^{2} } \right)\to \kappa \ \ \rm{for} \ \ \delta >0.
\end{eqnarray}
\end{subequations}
Now a strong coupling condition can be achieved at $\delta >0$. An appropriate laser field generation and amplification on transition ${\left| 2 \right\rangle} \to {\left| 1 \right\rangle} $ can be obtained in the presence of OCs only -- see Fig.\ref{fig1}(c). On the other hand, for $\delta <0$ one can obtain a vanishing coupling between atomic DS ensemble and cavity mode that corresponds to Raman-type lasing represented by the red arrow in Fig.\ref{fig1}(a), cf. \cite{24}.

In the following section we will focus mainly on the strong coupling between atomic DSs and cavity field when a new type of polaritons, that is, DS polaritons can be found.

\section{SUPERRADIANT PHASE TRANSITION}

Let us consider the processes characterized by \emph{negative} atom-light detuning $\delta <0$ and described by Hamiltonian \eqref{eq9_} under strong coupling condition
 \begin{equation} \label{eq29_}
\kappa_{12}\gg \max \left\{\gamma ,\Gamma _{c} ,\Gamma \right\},
\end{equation}
 where $\Gamma_{c}$ denotes cavity photon leakage, $\Gamma$ is spontaneous emission rate and $\gamma$ characterizes pressure broadening.

Strictly speaking, we suppose our system to be at full thermal equilibrium. The total excitation number $N_{\rm{ex}} =a^{\dag } a+\frac{1}{2} \sum _{j}S_{zj}^{\rm{DS}}  $ is a conserved quantity now. It is instructive to define polariton number density
\begin{equation} \label{eq13_}
\rho =\frac{1}{2} + \rho_{\rm{ex}} =\lambda ^{2} +\frac{1}{2} \left(\frac{1}{N_{\rm{at}} } \left\langle \sum _{j}S_{zj}^{\rm{DS}}  \right\rangle +1\right),
\end{equation}
that corresponds to the sum of an average number of cavity photons and atoms populating upper DS; $\rho_{\rm{ex}}=\frac{\left\langle N_{\rm{ex}} \right\rangle }{N_{\rm{at}} }$ is excitation number density. In \eqref{eq13_} $\lambda =\sqrt{\left\langle a^{\dag } a\right\rangle /N_{\rm{at}} } $ is normalized cavity field amplitude, which we take as an order parameter of the system.

Alternatively Eq.\eqref{eq13_} can be obtained by using DS polaritons. Actually, for describing macroscopic excitations of the effective two-level DS system one can define annihilation $\phi $ and creation $\phi ^{\dag } $ excitation operators by using transformation, cf. \cite{14},
\begin{subequations} \label{eq14_}
\begin{gather*}
\phi \simeq \sum _{j}^{N_{\rm{at}} }S_{-j}^{\rm{DS}} /\sqrt{N_{\rm{at}} }, \
\phi ^{\dag }  \simeq \sum _{j}^{N_{\rm{at}} } S_{+j}^{\rm{DS}} /\sqrt{N_{\rm{at}} }, \tag{\ref{eq14_}\rm{a,b}} \\
\sum _{j}^{N_{\rm{at}} }S_{zj}^{\rm{DS}} =2\phi ^{\dag } \phi -N_{\rm{at}}. \tag{\ref{eq14_}\rm{c}}
\end{gather*}
\end{subequations}
Here we restrict ourselves by the so-called low excitation density limit $\phi ^{\dag } \phi \ll N_{\rm{at}} $; excitation operators $\phi $, $\phi ^{\dag}$ possessing bosonic commutation relation $\left[ \phi,\phi^{\dag} \right] \simeq 1$, cf. \cite{21, 22}. Physically, a low density limit implies that it is the lower DS level $\left| 2 \right\rangle$ which is mainly populated, see (\ref{eq14_}c) and (\ref{eq8_}c); this being true under the condition of equilibrium Boltzmann distribution for DS level populations.

At the steady state for a coupled atom-light system Hamiltonian \eqref{eq9_} can be diagonalized with the help of operators \eqref{eq14_} by using unitary transformations
\begin{subequations} \label{eq15_}
\begin{equation*}
\Phi _{1} =Xa+C\phi,\ \
\Phi _{2} =X\phi -Ca, \eqno{(\ref{eq15_}\rm{a,b})}
\end{equation*}
\end{subequations}
where $X=\frac{1}{\sqrt{2} } \left(1+\frac{\Delta }{\sqrt{4\kappa_{12}^{2} +\Delta ^{2} } } \right)^{1/2} $, $C= \sqrt{1-X^2} $ are Hopfield coefficients, $\Delta =\delta _{c} -\Omega _{R} $ is effective detuning. Physically, $\Delta =\delta _{c} +\omega _{L} -(\Omega _{R} +\omega _{L} )=\omega _{c} -(\Omega _{R} +\omega _{L} )$ defines the detuning of a cavity field with frequency $\omega _{c} $ from blue Mollow triplet sideband frequency $\Omega _{R} +\omega _{L} $ -- see Fig.\ref{fig2}. We have $\Omega _{R} +\omega _{L} \approx \omega _{\rm{at}} $ for $\delta <0$ and $\Delta $ simply defines detuning of cavity field from atomic transition, that is $\Delta \simeq \omega _{c} -\omega _{\rm{at}} $.

Operators $\Phi _{1,2}$  in Eqs. (\ref{eq15_}a) and (\ref{eq15_}b) characterize upper and lower branch DS polaritons, which are mixed states of a cavity field and DS macroscopic polarization. Thus, the average total number of polaritons $N_{\rm{pol}} =\sum _{i=1,2}\left\langle \Phi _{i}^{\dag } \Phi _{i}  \right\rangle  $ normalized to atom number $N_{\rm{at}} $ is relevant to polariton number density $\rho$ in \eqref{eq13_}.

To determine the order parameter $\lambda $ at thermal equilibrium, we apply a variational (thermodynamic) approach, see e.g. \cite{30}. In this case, the partition function $Z=\mathrm{Tr} \left[\exp \left(-H'/k_{B} T\right)\right]$ should be used; $H'=H_{12} -\mu N_{\rm{ex}} $ is a modified Hamiltonian. The function $Z(T)$ describes a grand canonical ensemble with the finite (nonzero) chemical potential $\mu $. The estimate of the partition function $Z$ could be given in coherent state basis for the cavity photonic field, cf. \cite{31}.
Neglecting fluctuations of the optical field as well as atom-field correlations, one can obtain in the semiclassical limit
\begin{equation} \label{eq16_}
\tilde{\Omega }_{c} \lambda =\frac{\lambda \kappa_{12}^{2} \tanh \left[\hbar \Theta / 2 k_{B}T \right]}{ \Theta} ,
\end{equation}
where we made denotations $\tilde{\Omega }_{R} =\Omega _{R} -\mu $,  $\tilde{\Omega }_{c} =\delta _{c} -\mu $ and $\Theta \equiv \sqrt{\tilde{\Omega }_{R}^{2} +4\kappa_{12}^{2} \lambda ^{2} } $. Equation \eqref{eq16_} is a gap equation that characterizes the phase transition problem in different physical systems, see e.g. \cite{30,31,32}. In our case Eq. \eqref{eq16_} describes a second order phase transition to a superradiant phase characterized by $\lambda >0$.

By using \eqref{eq13_} and the partition function $Z(T)$ we can get an expression for  polariton density $\rho $ versus temperature:
\begin{equation} \label{eq17_}
\rho =\frac{1}{2} +\lambda ^{2} -\frac{\tilde{\Omega }_{R}  \tanh \left[\hbar \Theta / 2 k_{B}T \right]}{2 \Theta} .
\end{equation}
Combining \eqref{eq16_} and \eqref{eq17_} for $\mu $ parameter we have
\begin{equation} \label{eq18_}
\mu _{1,2} =\frac{1}{2} \left\{\delta _{c} +\Omega _{R} \pm \Omega _{ R\  {\rm eff}} \right\},
\end{equation}
where $\Omega _{R \  {\rm eff}} =\sqrt{\Delta ^{2} -8\kappa_{12}^{2} \left(\rho -\lambda ^{2} -1/2\right)} $. At low polariton densities and $\lambda =0$ Eq. \eqref{eq18_} defines the normal state for upper ($\mu _{1} $) and lower ($\mu _{2} $) polariton branch frequencies. Below we focus our attention on LB DS polaritons only, because at the full thermal equilibrium one can expect that a lower DS-polariton branch should be much more populated. In particular, in this case we can assume that $\tilde{\Omega }_{R} =\frac{1}{2} \left\{-\Delta +\Omega _{R \   {\rm eff}} \right\}$ and $\tilde{\Omega }_{c} =\frac{1}{2} \left\{\Delta +\Omega _{R \   {\rm eff}} \right\} $ if we use Eq. \eqref{eq18_}.

\begin{figure}
\includegraphics[scale=0.5]{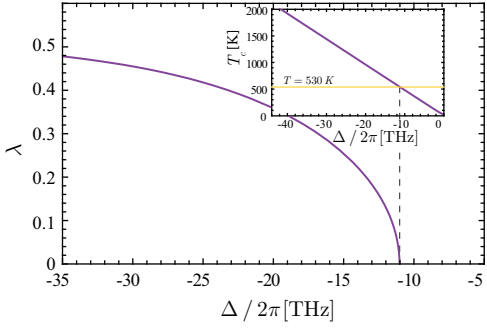}
\caption{\label{fig3}(Color online) The dependence of order parameter $\lambda $ on dressed atom-cavity field detuning $\Delta $ for a fixed value of DS polariton density $\rho =0.27$, corresponding to the system having the temperature $T=530$ K with $\kappa /2\pi =0.62$ THz, $\Omega /2\pi =1$ THz and $\delta /2\pi =-11$ THz. The dependence of critical temperature $T_{C} $  on detuning $\Delta $ for the same value of $\rho $ and $T$ is plotted in the inset .}
\end{figure}

In the limit of non-resonant atom-field interaction for OC processes it is instructive to suppose that detuning $|\Delta |\sim |\delta | \gg \kappa_{12}$. The critical temperature $T_{C} $ of a phase transition (for given atom-field detuning $\delta $) can be obtained from \eqref{eq16_} for $\lambda =0$ and looks like
\begin{equation} \label{eq20_}
T_{C} \approx \frac{\hbar \Delta }{2k_{B} \tanh ^{-1} \left[2\rho -1\right]}.
\end{equation}
With the help of Eq. \eqref{eq17_} it is possible to obtain the expression for the order parameter $\lambda $ that obeys
\begin{equation} \label{eq22_}
\lambda \approx \lambda _{\infty } \left\{1-\frac{1}{\rho \left(1+\left(\rho ^{-1} -1\right)^{\zeta /\zeta _{c} } \right)} \right\}^{1/2} ,
\end{equation}
where $\zeta =\hbar \Delta /k_{B} T$ is a vital parameter for our problem, $\lambda _{\infty } $ is an order parameter in ``zero temperature'' limit $\hbar \tilde{\Omega }_{R} \gg k_{B} T$; $\zeta _{c} =-\ln \left[\rho ^{-1} -1\right]$ is a critical value of parameter $\zeta $ in the limit when $\left|\Delta \right| \gg \kappa_{12}$ that defines a phase boundary between normal and superradiant states.
The dependence of order parameter $\lambda $ and critical temperature $T_{C} $ of the phase transition (inset)  on atom-cavity field detuning $\Delta $  is presented in Fig.\ref{fig3} for  a fixed value of excitation number density $\rho $. When the critical temperature $T_{C} $ exceeds the temperature of the system (horizontal solid (yellow) line) a non-zero value of coherent light amplitude $\lambda $ is obtained.

For large, negative detunings $\Delta $ the phase transition temperature is high enough even for low polariton number densities $\rho $ due to the photon-like character of LB polaritons ($\Phi _{2} \approx -a$). Our special interest is in the case when $\Delta \approx -\Omega _{R} $ that corresponds to the degenerate frequency of a cavity mode $\omega _{c} \approx \omega _{L} $ ($\delta _{c} =0$). In fact, in this case we obtain from \eqref{eq20_} the same result as is derived from \cite{14} for a superradiant high temperature phase transition without cavity. However, in the present case such results can be also interpreted as a phase transition to BEC state for DS polaritons if we take into account the cavity geometry and trapping potential, cf. \cite{20, 32}.

On the other hand, for vanishing detunings $\Delta $ the critical temperature diminishes. In particular, for half photon half matter-like LB polaritons that correspond to the resonant case of $\Delta =0$ ($\omega _{c} \approx \omega _{\rm{at}} $ -- see Fig. \ref{fig2}(a)) the temperature \eqref{eq20_} of a phase transition becomes essentially lower.

\section{DS LASING FOR ${\left| 1(N) \right\rangle} \to {\left| 2(N-1) \right\rangle} $ TRANSITION }

In the real experiment a thermodynamically full  equilibrium cannot be achieved because of the spontaneous emission and finite lifetime of photons in the cavity. The aim of this section is to create the conditions for crossover from superradiant phase transition to lasing.

We start from the master equation for density matrix $\sigma $ in the presence of OCs, radiative (spontaneous) relaxation and cavity leakage processes. This can be written as follows:
\begin{equation} \label{eq23_}
\dot{\sigma }=-\frac{i}{\hbar } \left[H,\sigma \right]+L_{\rm{rad}} \sigma +L_{\rm{coll}} \sigma + L_{\rm{cav}}\sigma,
\end{equation}
where the last three terms
\begin{subequations} \label{eq24_}
\begin{equation}
 L_{\rm{rad}} \sigma =\Gamma \sum _{j}^{N_{\rm{at}} }\left\{s_{-j} \sigma s_{+j} -\frac{1}{2} \left(s_{+j} s_{-j} \sigma +\sigma s_{+j} s_{-j} \right)\right\},
 \end{equation}
 \begin{equation}
L_{\rm{coll}} \sigma =\sum _{j}^{N_{\rm{at}} }\left\{-\frac{\gamma }{2} \sigma +2\gamma s_{zj} \sigma s_{zj} -i\eta \left[s_{zj} ,\sigma \right]\right\},
\end{equation}
\begin{equation}
L_{\rm{cav}} \sigma = \Gamma_{c}\left( 2a \sigma a^{\dag} - a^{\dag}a \sigma - \sigma a^{\dag} a \right),
\end{equation}
\end{subequations}

\noindent account for the spontaneous emission (\ref{eq24_}a), collisions with buffer gas atoms (\ref{eq24_}b) and a leakage of photons out of the cavity (\ref{eq24_}c). In (\ref{eq24_}b) $\gamma$ is collisional broadening, $\eta$ is collisional phase shift. In Eq. \eqref{eq23_} we suppose that Hamiltonian $H$ is given by Eq. \eqref{eq9_}.

Neglecting quantum atom-field correlations and inhomogeneous broadening that is significantly smaller than the collisional one we can obtain  Maxwell-Bloch-like equations
\begin{subequations} \label{eq25_}
\begin{gather}
\dot{\lambda }=-(i\delta _{c} +\Gamma _{c} )\lambda -i\kappa_{12}S, \\
\dot{S}=-\left(i\left(\Omega _{R} +\eta _{1} \right)+\Gamma _{1} +\gamma _{1} \right)S+i\kappa_{12}\lambda S_{z}, \\
\dot{S}_{z}  =-2w\left(S_{z} -S_{z}^{(\rm{eq})} \right)-\Gamma _{+} S_z+\Gamma_{-} +2i\kappa_{12}(S\lambda ^{*} -S^{*} \lambda )
\end{gather}
\end{subequations}
for new variables which are a normalized coherent amplitude of cavity field $\lambda$, collective atomic excitation ($S$) and population imbalance ($S_{z} $); the latter two are defined as
\begin{subequations} \label{eq26_}
\begin{gather}
S =\frac{1}{N_{\rm{at}} } \sum _{j=1}^{N_{\rm{at}}}\sum _{N=0}^{\infty }{\left\langle 1(N) \right|} _{j} \sigma {\left| 2(N) \right\rangle}_{j} ,\\
S_{z} =\frac{1}{N_{\rm{at}} } \sum _{j}^{N_{\rm{at}} }\sum _{N=0}^{\infty }\left({\left\langle 1(N) \right|} _{j} \sigma {\left| 1(N) \right\rangle}_{j} -{\left\langle 2(N) \right|} _{j} \sigma {\left| 2(N) \right\rangle}_{j} \right).
\end{gather}
\end{subequations}

\noindent In Eqs. \eqref{eq25_} the following notations are introduced:
\begin{gather}
\ w=\frac{\gamma \mathop{\sin }\nolimits^{2} (2\theta )}{2}, \nonumber
\\
\Gamma _{\pm } =\Gamma \left(\sin ^{4} (\theta )\pm \cos ^{4} (\theta )\right),
\gamma _{1} =\gamma (\cos ^{4} \theta +\sin ^{4} \theta ), \nonumber
\\
\eta _{1} =\eta \cos 2\theta, \  \Gamma _{1} =\frac{\Gamma }{4} \left(2+\sin ^{2} 2\theta \right). \label{eq27_}
\end{gather}
Thereafter we assume that Rabi frequency $\Omega _{R} $ also includes phase shift $\eta _{1} $ introduced by the collisions, i.e. $\Omega _{R} +\eta _{1} \to \Omega _{R} $.

In \eqref{eq25_}  we have incorporated $S_z^{(\rm{eq})}=-\tanh\left[{\frac{\hbar \Omega_{R}}{2k_B T}}\right]$, that characterizes a thermodynamically equilibrium value of DS population imbalance, cf. \cite{Leg}. In the absence of a cavity (at $\lambda=0$) DS population imbalance $S_z$ relaxes toward its stationary value $S_{z}^{(\rm{st})} =\frac{2wS_{z}^{(\rm{eq})} +\Gamma _{-} }{2w+\Gamma _{+} }$. In the presence of thermalized OCs the difference of DS populations approaches it equilibrium value with the rate $2w$, i.e. $S_{z}^{(\rm{st})} \simeq S_{z}^{(\rm{eq})}$.

\begin{figure}
\includegraphics[scale=0.4]{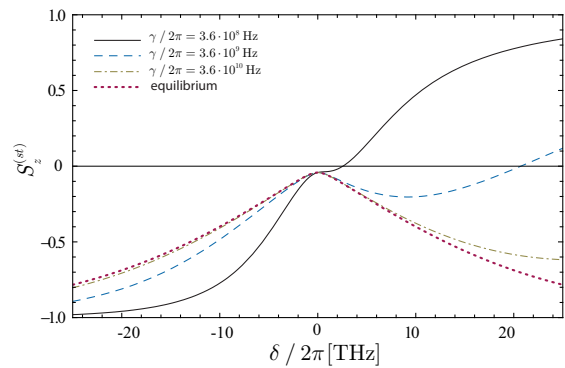}
\caption{\label{fig4}Dependence of a stationary value of DS population imbalance $S_{z} $ as a function of detuning $\delta /2\pi $ for different values of collisional broadening. Parameters are: $\Omega /2\pi $ = 1 THz, $\kappa /2\pi =0.62$ THz,  $\Gamma \simeq 37$ MHz.}
\end{figure}

In Fig.\ref{fig4} we represent the dependences for DS population imbalance $S_{z}  $ as a function of detuning $\delta $. A thermodynamically full equilibrium behavior of $S_{z} $ is shown by the dotted (red) curve in Fig.\ref{fig4}. From Fig. \ref{fig4} it is clearly seen that a  thermodynamically equilibrium state could be achieved at the far red detuned tails ($\delta <0$) of DS population imbalance $S_{z}  $  under high enough buffer gas pressures which correspond to large values of $\gamma $.

The terms containing $\Gamma$ in \eqref{eq25_} characterize the influence of spontaneous emission on the process of thermalization. To minimize this effect it is necessary to require that the rate of thermalization of atom-field DS  be much higher than the effective rate of the spontaneous emission. In other words thermalization occurs when condition $w \gg \Gamma_{+}$ (see (\ref{eq25_}c)) is fulfilled or, in perturbative limit \eqref{eq3_}, inequalities
\begin{equation} \label{eq28_}
\frac{\Gamma }{\gamma } \ll \frac{\Omega^{2} }{\delta ^{2} } \ll 1
\end{equation}
are satisfied -- see \cite{19}.

The suppression of DS thermalization process due to the spontaneous emission leads to the formation of thermodynamically quasi-equilibrium (dashed-dotted curve in Fig.\ref{fig4}) or completely non-equilibrium (dashed (blue) and solid (black) curves, respectively) of the coupled atom-light states for which the condition \eqref{eq28_} is violated.

For highly non-equilibrium coupled atom-light states $S_{z}  \approx 1$ (solid curve in Fig.\ref{fig4}) at positive detuning $\delta > 0$ we obtain the inversion for DS population, that corresponds to the inversionless two-level atomic system. In this case the role of pressure broadening (parameter $\gamma $) is responsible for the rapid dephasing of DS (see (\ref{eq25_}b)) at the rate $\gamma_{1}$.

Below we focus on the thermodynamically non-equilibrium limit for a coupled atom-light system that can also be  connected with a large cavity field dissipation rate $\Gamma _{c} $, as an example. Stationary solutions $\lambda =\lambda e^{-i\mu t} $, $S=Se^{-i\mu t} $ and $\dot{S}_{z}^{} =0$  of Eqs. (\ref{eq25_}a,b) yield to
\begin{multline} \label{eq32_}
\mu _{1,2} =\frac{1}{2} \left\{\left(\delta _{c} -i\Gamma _{c} \right)+\left(\Omega _{R} -i(\Gamma _{1} +\gamma _{1} )\right)\pm \right. \\
\left. \sqrt{\left(\left(\delta _{c} -i\Gamma _{c} \right)-\left(\Omega _{R} -i(\Gamma _{1} +\gamma _{1} )\right)\right)^{2} -4\kappa_{12}^{2} \bar{S}_{z} } \right\},
\end{multline}
that implies relevant ``chemical potential'' for polaritons. The imaginary part of Eqs. \eqref{eq32_} suppresses coherent effects for cavity field $\lambda $ and is responsible for the dissipation of polariton states, when condition \eqref{eq29_} is satisfied, cf. \cite{22}. It is justified to choose $\bar{S}_{z} $ for lower branch eigenstates such as $\mathrm{Im}(\mu _{2} )=0$. In this case one can obtain from \eqref{eq32_} a threshold value $S_{z}^{(\rm{thr})}$ of the stationary population inversion level $\bar{S}_{z} $:
\begin{equation} \label{eq33_}
S_{z}^{(\rm{thr})} =\frac{\Gamma _{c} \left(\Gamma _{1} +\gamma _{1} \right)}{\kappa_{12}^{2} } \left(1+\frac{\Delta ^{2} }{\left(\Gamma _{c} +\Gamma _{1} +\gamma _{1} \right)^{2} } \right).
\end{equation}
At the same time the real part of ``chemical potential'', that is $\mu \equiv \mathrm{Re}(\mu _{2} )$ takes the form
\begin{equation} \label{eq34_}
\mu =\delta _{c} -\frac{\Delta \Gamma _{c} }{\Gamma _{1} +\gamma _{1} +\Gamma _{c} } .
\end{equation}

Equation \eqref{eq34_} determines the characteristic frequency of a laser field generation under the condition \eqref{eq33_}.  Physically, $\mu $ signifies a relative frequency accounted from frequency $\omega _{L} $ of the pump light. Relation $\gamma _{1} \gg \Gamma _{1} ,\Gamma _{c} $  can be achieved for high pressure atomic vapor and large atom-light detunings $\delta$, when it is realized simultaneously with inequality $w<\Gamma$; this being done for gaining DS population inversion. Thus, Eq.\eqref{eq32_} results in $\mu \approx \delta _{c} $.

The stationary population of DSs can be found out by solving the set of Eqs. \eqref{eq25_} that yields to
\begin{equation} \label{eq34b_}
\bar{S}_{z} =\frac{S_{z}^{(\rm{st})} \left|\tilde{\Omega }_{R,\rm{eff}}  \right|^{2} }{\left|\tilde{\Omega }_{R,\rm{eff}}  \right|^{2} +\frac{4(\Gamma _{1} +\gamma _{1} )}{2w+\Gamma _{+} } \kappa_{12}^{2} \left|\lambda \right|^{2} }
\end{equation}

\noindent where we define $\tilde{\Omega }_{R,\rm{eff}}  =\frac{1}{2} \left\{-\Delta -i\Gamma _{\rm{eff}} +\sqrt{\left(\Delta +i\Gamma _{\rm{eff}} \right)^{2} -4\kappa_{12}^{2} \bar{S}_{z} } \right\}$, and $\Gamma _{\rm{eff}} =\Gamma _{1} +\gamma _{1} -\Gamma _{c} $, cf. \eqref{eq16_}.

To specify the properties of the order parameter $\lambda$ in this case one can represent the polariton number density \eqref{eq13_} that evolves in time according to
\begin{equation} \label{eq35_}
\dot{\rho }=-0.5(2w+\Gamma _{+} )(S_{z} -S_{z}^{(\rm{st})} )-2\Gamma _{c} \left|\lambda \right|^{2} \end{equation}
in the presence of dissipation/dephasing effects.
At the steady state we can put in \eqref{eq35_} $\dot{\rho }=0$ and $S_{z} =S_{z}^{(\rm{thr})} $, that immediately leads to the equation for   $\left|\lambda \right|^{2} $ in the form:
\begin{equation} \label{eq36_}
\left|\lambda \right|^{2} =\frac{\left(2w+\Gamma _{+} \right)}{4\Gamma _{c} } \left(S_{z}^{(\rm{st})} -S_{z}^{(\rm{thr})} \right).
\end{equation}

In the DS laser theory Eq. \eqref{eq36_} plays the same role as Eq. \eqref{eq16_} in the theory of thermodynamically equilibrium phase transition to superradiant state with a cavity photonic field, c.f. \cite{31}. Actually, Eq. \eqref{eq36_} indicates the region, where a cavity field has nonzero value. Lasing can occur only for DS population imbalance $S_{z}^{(\rm{st})} >0$ obeying the condition $S_{z}^{(\rm{st})} >S_{z}^{(\rm{thr})} \ge 0$ that  corresponds  to the non-equilibrium phase transition for atom-light system taken for positive $\delta $ -- see solid curve in Fig. \ref{fig4}.

It is worth emphasizing that the threshold of DS population imbalance $S_{z,\min }^{(\rm{thr})} ={\Gamma _{c} \left(\Gamma _{1} +\gamma _{1} \right)}/{\kappa_{12}^{2} } $  is minimal for a cavity light detuning which satisfies resonant condition $\Delta =0$. In Fig.\ref{fig2}(b) we have shown relevant transition frequencies for a laser field generation taking place at  $\delta >0$ under the perturbative limit \eqref{eq3_}. It is easy to see that the resonant case $\Delta =0$ can be realized when the frequency of cavity mode $\omega _{c} $ is in the vicinity of the blue component of Mollow triplet that corresponds to transition ${\left| 1 \right\rangle} \to {\left| 2 \right\rangle} $ in the DS level picture -- see Fig.\ref{fig1}.  On the other hand, $S_{z}^{(\rm{thr})} $ increases with detuning $\Delta \ne 0$ from the blue Mollow component. For example, at $\Delta =-\Omega _{R} $ we have $S_{z}^{(\rm{thr})} \approx \frac{\Gamma _{c} \delta ^{2} }{\kappa_{12}^{2} \left(\Gamma _{c} +\Gamma _{1} +\gamma _{1} \right)} \gg S_{z,\min }^{(\rm{thr})} $.

Noticing that vanishing threshold DS population imbalance $S_{z}^{(\rm{thr})} \simeq 0$ can be obtained under the inequality
\begin{equation} \label{eq37_}
\kappa_{12} \gg \sqrt{\gamma \Gamma _{c} } ,
\end{equation}
which can also be associated with some modification of the effective strong coupling condition \eqref{eq29_}. In fact, in this case we deal with the threshold-less laser field generation for which the order parameter is maximal and is equal to $\left|\lambda \right|\approx \sqrt{{\Gamma S_{z}^{(\rm{st})} }/{4\Gamma _{c} } } $.

\begin{figure*}
\includegraphics[scale=0.35]{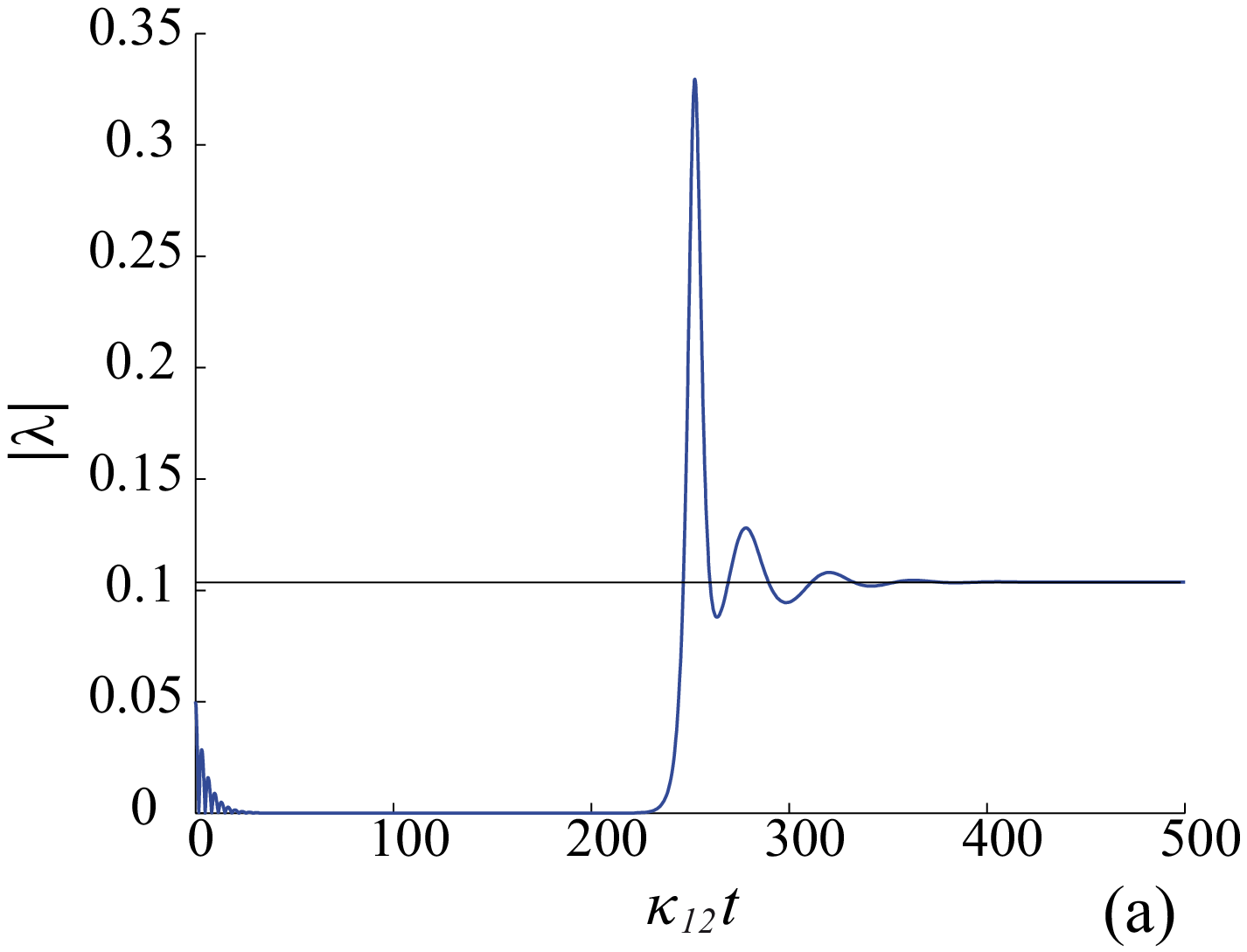}
\includegraphics[scale=0.35]{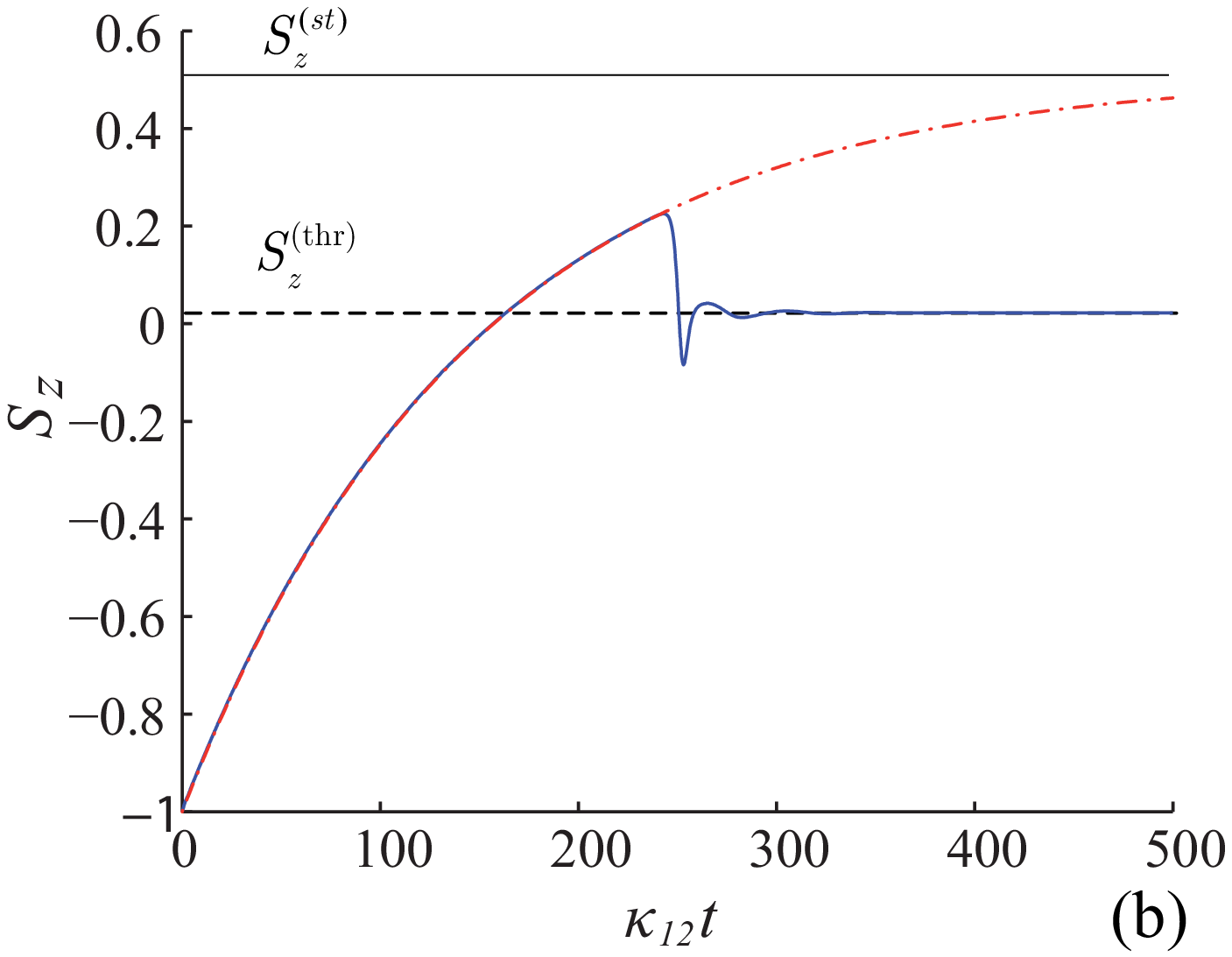}
\includegraphics[scale=0.35]{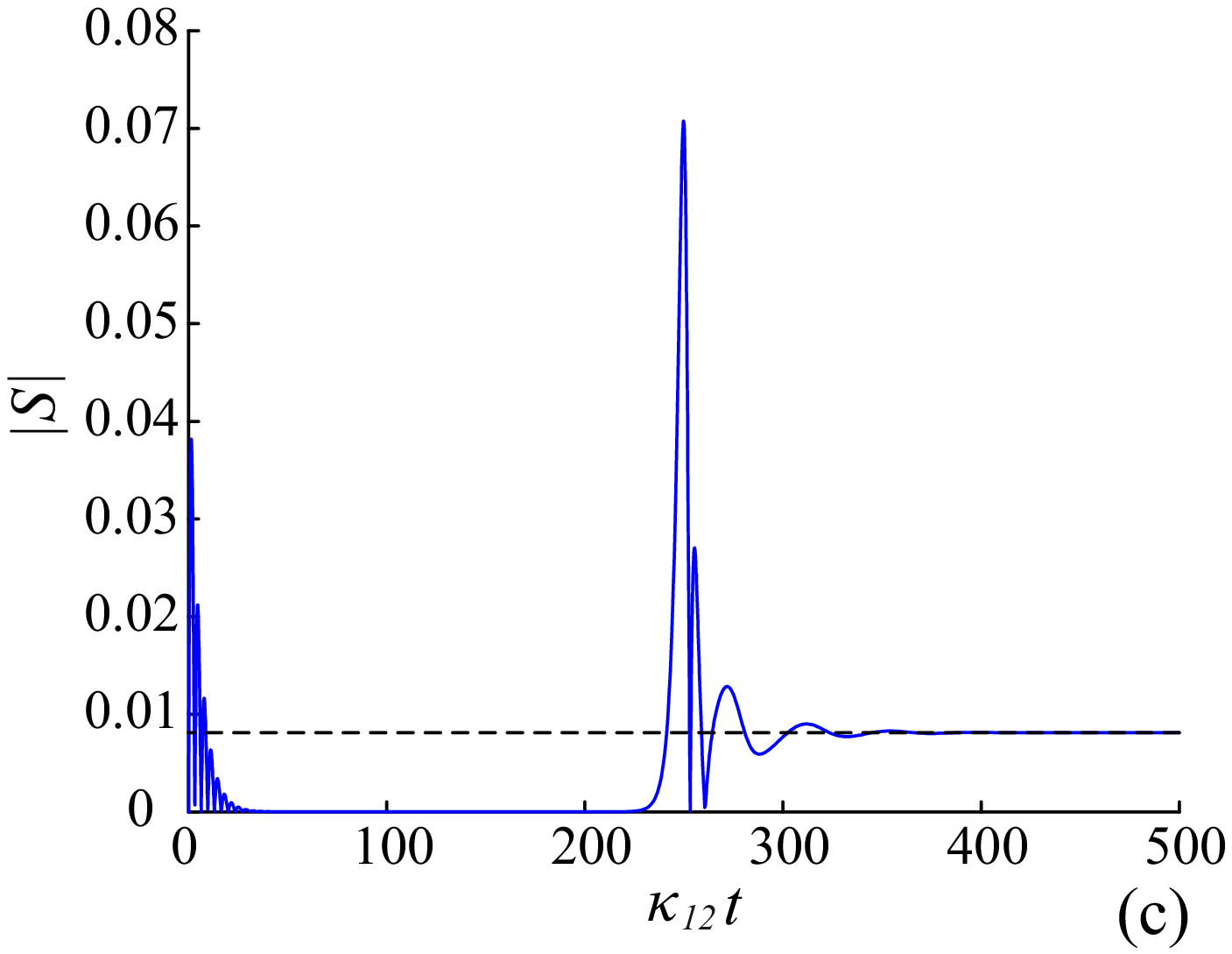}
\caption{\label{fig6}(Color online) Dependencies of (a) cavity field amplitude $\left|\lambda \right| $, (b) DS  population imbalance $S_{z}  $, and  (c) DS polarization  $\left|S\right|$ as a function of reduced time $\kappa_{12}t$. Parameters are: $\Omega /2\pi $ = 1 THz, $\kappa /2\pi =0.62$ THz, $\delta _{c} =\Omega _{R} $, $\Gamma _{c} /2\pi =100$~MHz, $\gamma /2\pi =0.36$ GHz, $\delta /2\pi =11$ THz. Initial conditions are: $\rm{Re}\left[\lambda (0)\right]=0.05$, $\rm{Im}\left[\lambda (0)\right]=0$, $\rm{Re}\left[S(0)\right]=\rm{Im}\left[S(0)\right]=0$, $S_{z} =-1$. The dot-dashed (red) curve on (b) describes the behavior of the system without cavity, cf. \cite{19}.}
\end{figure*}

In Fig. \ref{fig6} we represent the numerical solution of a full set of Eqs. \eqref{eq25_} revealing  non-equilibrium dynamics of modulus of the order parameter (cavity field amplitude) $\left|\lambda \right|$, DS population imbalance ($S_{z}  $) and DS polarization ($\left|S\right|$) in the presence of OCs. We suppose that all atoms initially occupy the lower DS level ${\left| 2(N) \right\rangle} $; that is, $S_{z} (t=0)=-1$ with zero value of DS polarization $S(t=0)=0$. Without a cavity, i.e. at $\lambda =0$, a coupled atom-light system exhibits the behavior with $\bar{S}_{z} =S_{z}^{(\rm{st})} $ that is inherent to the transient quasi-equilibrium dynamics described in \cite{19}. Lasing occurs for $S_{z} (t=\tau _{L} )=S_{z}^{(\rm{thr})} $ at  characteristic time scales  $\tau _{L} $ which  can be taken from the evolution of DS population imbalance in the absence of the cavity field $S_{z} =S_{z}^{(\rm{st})} +(S_{z}(t=0) -S_{z}^{(\rm{st})} )e^{-(2w+\Gamma _{+} )t} $. Fig. \ref{fig6}(b) and Fig. \ref{fig6}(c) indicate rapidly vanishing temporal  oscillations around the value $S_{z}^{(\rm{thr})} $ for DS population imbalance $S_{z} $ and modulus of DS polarization  $\left|S\right|$ at  $t>\tau _{L} $. For the plots in Fig. \ref{fig6} we  have  $\tau _{L} \approx 20$ ns.   Noticing that the estimated  time of thermalization  $T_{therm} ={2\pi \delta ^{2} }/{\gamma \Omega^{2} } $ (see \cite{19}) for the parameters given in Fig. \ref{fig6} is large enough   ($T_{therm} \simeq 0.34\rm{\mu s}$)  as compared to $\tau _{L} $ and characteristic times separated two successive acts of collisions  and spontaneous decay from the upper level.

\begin{figure}
\includegraphics[scale=0.45]{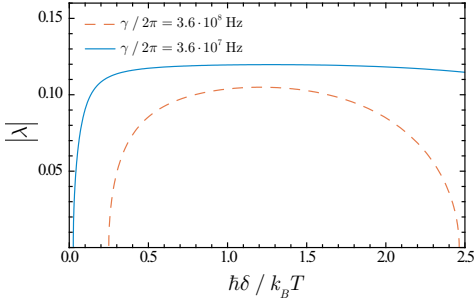}
\caption{\label{fig7} (Color online) Modulus of order parameter $\left|\lambda \right|$ vs reduced positive atom-light detuning $\delta $ for $T= 530$ K temperature of atomic gas confined in the cavity. The parameters are $\Omega /2\pi $ = 1 THz, $\kappa /2\pi =0.62$ THz, $\Gamma _{c} /2\pi =100$~MHz, $\Gamma \simeq 37$ MHz, $\Delta =0$.}
\end{figure}

In Fig.\ref{fig7} we give the dependencies of modulus of the order parameter $\left|\lambda \right|$ as a function of \emph{positive} atom-light detuning $\delta $ taken for transition ${\left| 1 \right\rangle} \to {\left| 2 \right\rangle} $ for $S_{z}^{(\rm{thr})} =S_{z,\min }^{(\rm{thr})} $ and for different values of collisional broadening.  DS lasing occurs within the domain where the order parameter $\left|\lambda \right|\ne 0$. In this sense we can speak about the analogy between lasing and thermodynamically equilibrium phase transition to superradiant phase, cf. Fig.\ref{fig3}.  Noticing that for moderate values of detuning $\delta $ the value of a laser field amplitude grows due to the increase of the steady state DS population imbalance $S_{z}^{(\rm{st})} $ -- see Fig.\ref{fig4}. However, for a large enough $\delta $ the magnitude of $S_{z}^{(\rm{st})} $ is saturated while the effective coupling parameter $\kappa_{12}$ vanishes according to $1/\delta ^{2} $ (see (\ref{eq11_}b)) and a strong coupling condition \eqref{eq37_} is broken; the order parameter vanishes as well.

\begin{figure}
\includegraphics[scale=0.45]{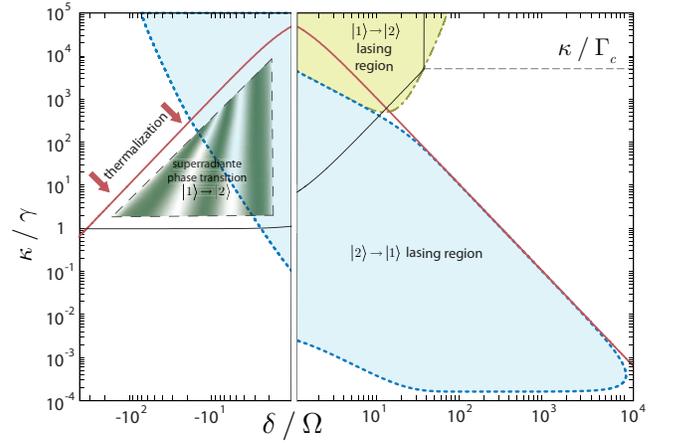}
\caption{\label{fig8} (Color online) Phase diagram. The parameters are  $\Omega /2\pi $ = 1 THz, $\kappa /2\pi =0.62$ THz, $\Gamma _{c} /2\pi =100$~MHz. The detuning $\Delta $ for each transition was chosen so as to satisfy the most favorable condition for generation, i.e. $\Delta =0$ for ${\left| 1 \right\rangle} \to {\left| 2 \right\rangle} $, and $\Delta =-2\Omega _{R} $ for ${\left| 2 \right\rangle} \to {\left| 1 \right\rangle} $ transitions, respectively. Red (solid) lines define boundaries for DS thermalization condition \eqref{eq28_}. Black (solid) curve corresponds to condition $\kappa_{12} = \mathrm{max} \{ \gamma,\Gamma,\Gamma_{c} \}$ defining the boundary of a strong coupling limit \eqref{eq29_}. Dashed (green) triangle domain corresponds to the proposed equilibrium superradiant phase transition. White areas correspond to the order parameter $\lambda =0$.}
\end{figure}

In Fig.\ref{fig8} we represent a phase diagram exhibiting the dependence of ratio $\kappa /\gamma $ (for fixed $\kappa $) on normalized atom-light detuning $\delta /\Omega $ (for fixed $\Omega $). Thermalization of atomic DSs requires a large enough collisional broadening parameter $\gamma $. However, at the same time we require a strong coupling between the effective DS atomic system and the cavity field that implies the enhancement of ratio $\kappa /\gamma $. For negative detuning $\delta $ conditions \eqref{eq28_}, \eqref{eq29_} can be fulfilled simultaneously in the small (green) area which is far from resonance and corresponds to the superradiant state where thermodynamical approach is justified.

Crossover to the lasing that corresponds to transition ${\left| 1 \right\rangle} \to {\left| 2 \right\rangle} $ in DS basis occurs at the positive detuning $\delta $ - dark area in Fig.\ref{fig8}. It is interesting to note that the width of the area where $\left|\lambda \right|\ne 0$ depends essentially on collisional broadening parameter $\gamma $, see also Fig.\ref{fig7}.  This area becomes smaller with the increase of $\gamma $. In this sense collisions lead to the suppression of DS laser field generation.

\section{DS LASING FOR ${\left| 2(N) \right\rangle} \to {\left| 1(N-1) \right\rangle} $ TRANSITION}

Now we pay our attention to the transition ${\left| 2 \right\rangle} \to {\left| 1 \right\rangle} $ in DS manifold picture (see Fig.\ref{fig1}), described by Hamiltonian \eqref{eq10_}. Proceeding as in the previous section it is possible to get Maxwell-Bloch-like equations for normalized cavity field $\lambda $, DS polarization $S$ and DS population imbalance $S_{z} $ in the form
\begin{subequations} \label{eq39_}
\begin{gather}
\dot{\lambda }=-\left(i\delta _{c} +\Gamma _{c} \right)\lambda +i\kappa_{21}S^{*},\\
\dot{S}=-\left[i\Omega _{R} +\Gamma _{1} +\gamma _{1} \right]S-i\kappa_{21}\lambda ^{*} S_{z} , \\
\dot{S}_{z}  =-\left(2w+\Gamma _{+} \right)\left(S_{z} -S_{z}^{(\rm{st})} \right)+2i\kappa_{21}(S^{*} \lambda ^{*} -S\lambda ).
\end{gather}
\end{subequations}

Firstly, we examine Eqs. \eqref{eq39_} under the strong atom-cavity field condition when detuning $\delta $ is positive, see (\ref{eq12_}b). Stationary solutions $\lambda =\lambda e^{-i\mu t} $and $S^{*} =S^{*} e^{-i\mu t} $ of Eqs.\eqref{eq39_} lead to
\begin{multline} \label{eq40_}
\mu _{1,2} =\frac{1}{2} \left\{\left(\delta _{c} -i\Gamma _{c} \right)-\left(\Omega _{R} +i(\Gamma _{1} +\gamma _{1} )\right)\pm \right. \\
\left. \sqrt{\left(\left(\delta _{c} -i\Gamma _{c} \right)+\left(\Omega _{R} +i(\Gamma _{1} +\gamma _{1} )\right)\right)^{2} +4\kappa_{21}^{2} \bar{S}_{z} } \right\}.
\end{multline}

Proceeding as in the previous section for ``chemical potential'' which is a relative frequency of laser field generation one can obtain $\mu =\delta _{c} -\frac{\left(\Delta +2\Omega _{R} \right)\Gamma _{c} }{\Gamma _{1} +\gamma _{1} +\Gamma _{c} } $. The generation itself, is determined by the condition $S_{z}^{(\rm{st})} \le S_{z}^{(\rm{thr})} $, where
\begin{equation} \label{eq41_}
S_{z}^{(\rm{thr})} =-\frac{\Gamma _{c} \left(\Gamma _{1} +\gamma _{1} \right)}{\kappa_{21}^{2} } \left(1+\left(\frac{\Delta +2\Omega _{R} }{\Gamma _{c} +\Gamma _{1} +\gamma _{1} } \right)^{2} \right)
\end{equation}
is a threshold population imbalance for DSs which is always negative.

\begin{figure}
\includegraphics[scale=0.5]{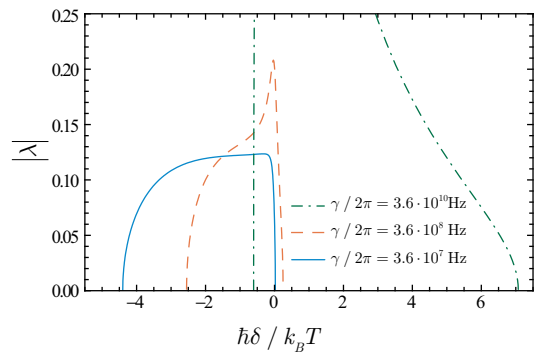}
\caption{\label{fig9} (Color online) Modulus of order parameter $\left|\lambda \right|$ vs reduced atom-light detuning  $\delta $ for  transition ${\left| 2(N) \right\rangle} \to {\left| 1(N-1) \right\rangle} $. The parameters are: $T = 530$ K, $\Omega /2\pi $ = 1 THz, $\kappa /2\pi =0.62$ THz, $\Gamma _{c} /2\pi =100$~MHz, $\Delta =-2\Omega _{R} $.}
\end{figure}

Noticing that stationary DS population imbalance  $\bar{S}_{z} $ is characterized by the same equation as (\ref{eq34b_}) but with polarization frequency  $\mu $ determined in \eqref{eq40_}. In this case the modulus square of the order parameter is determined as
\begin{equation} \label{eq42_}
\left|\lambda \right|^{2} =\frac{\left(2w+\Gamma _{+} \right)}{4\Gamma _{c} } \left(S_{z}^{(\rm{thr})} -S_{z}^{(\rm{st})} \right).
\end{equation}
Since threshold value $S_{z}^{(\rm{thr})} $ of DS population imbalance is always negative (see \eqref{eq41_}) and cf. \eqref{eq33_}) we expect that  $S_{z}^{(\rm{st})} \le S_{z}^{(\rm{thr})} <0$, particularly for obtaining lasing in DS system, i.e. for  $\left|\lambda \right|^{2} >0$.  It is important that lasing can be obtained both in positive ($\delta >0$) and negative ($\delta <0$) domains, see Fig.\ref{fig9} and blue areas in Fig.\ref{fig8}.  This situation radically differs from lasing conditions examined in the previous section where lasing occurs only at $\delta >0$.  Actually, in the system being far from thermodynamic equilibrium, one can obtain $S_{z}^{(\rm{st})} <0$ for $\delta >0$ just in the vicinity of atom-field resonance -- see solid curve in Fig.\ref{fig4}. Meanwhile, true atomic population inversion characterized by DS population   \emph{without inversion} occurs under the condition \eqref{eq28_} and happens for large enough $\delta $ and relatively large collisional broadening $\gamma $ -- see Fig.\ref{fig1}(c) and dashed-dotted line in Fig.\ref{fig4}.

In Fig.\ref{fig9}  dashed and dashed-dotted curves clearly show the enhancement of the order parameter $\left|\lambda \right|$  by increasing parameter $\gamma $.  The effective coupling parameter $\kappa_{21}$ is maximal and equal to  $\kappa_{21}\approx \kappa $ at $\delta > 0$, see (\ref{eq12_}b).  From Eqs. \eqref{eq41_}, \eqref{eq42_} it is  obvious that the lowest threshold level $S_z ^{(\rm{thr})}$ can be achieved with  $\Delta =-2\Omega _{R} $, i.e. when the frequency of a cavity mode is tuned to the red Mollow triplet (see Fig. \ref{fig2}), and we have $S_{z}^{(\rm{thr})} \simeq 0$ due to the fulfillment  of a strong atom-field coupling condition $\kappa_{21} >\sqrt{\gamma \Gamma _{c} } $, cf. \eqref{eq37_}.

Thus, one can conclude that for $\delta >0$ lasing phenomenon that is characterized by a wide area in Fig. \ref{fig8}  occurs in the presence of  quasi-thermalized DSs, cf. \cite{18}.

\section{CONCLUSION}

Let us briefly summarize the results obtained. We have considered the problem of both thermodynamically  equilibrium and non-equilibrium phase transitions with  coupled atom-light states, i.e. dressed states.  The essence of our description is the so-called  OCs occurring with the emission or absorption of two-level (rubidium) atoms placed in the cavity in the presence of the collision with buffer gas particles. In this framework we analyze a rich picture of coherent effects that take place under the atomic transition within DS manifolds for a coupled atom-light system. In particular, thermodynamically equilibrium phase transition to some superradiant (coherent) phase becomes possible for the cavity field in an inversionless atomic system at the negative detuning of a pump field from atomic transition due to the establishment of a strong coupling regime and thermalization of atomic DS population simultaneously. Physically such a transition occurring in the cavity at high enough (500 K and above) temperatures can be explained in the terms of the phase transition of photon-like  DS polaritons  to some condensed (and/or superfluid) state in the cavity with a special geometry, cf. \cite{20}.

The suppression of DS thermalization process due to spontaneous emission leads to the formation of thermodynamically quasi-equilibrium or completely non-equilibrium  coupled atom-light states. In this case the atomic system in the cavity exhibits non-equilibrium phase transition to lasing in  DS basis under the DS population inversion  only. Noticing that the vanishing threshold for DS population imbalance $S^{(\mathrm{thr})}\simeq 0$  can be achieved under the fulfillment of the inequality that represents some modification of a strong coupling condition.  The order parameter that is nothing else but the average number of photons in the cavity field in this limit depends on the ratio of atomic spontaneous emission rate to the cavity decay rate. It is important to emphasize that the lasing phenomenon  characterized by DS population imbalance  occurs at $\delta > 0$ in the presence of DS thermalization too, that, perhaps, contradicts the imperative of lasers as entirely non-equilibrium devices.

Hopefully, the theory developed in the paper can be also useful for describing some other systems where DS picture is of great importance. Practically, here we refer to the problem of lasing and amplification in the system of superconducting flux qubits coupled to a resonator in the presence of a strong driving field, cf. \cite{Ast}.

\section*{ACKNOWLEDGMENTS}
This work was supported by RFBR Grants No. 11-02-97513, 12-02-97529, 12-02-90419, 12-02-31601 and by Russian Ministry of Education and Science under Contract No. 3088.2012.2 and Federal program ``The development of a higher school scientific potential'' No. 2.4053.2011.

\end{document}